\documentclass{aa}  

\usepackage{graphicx}
\usepackage[english]{babel}
\usepackage{hyperref}
\usepackage{txfonts}
\usepackage{multirow}
\usepackage{multicol}

\newcommand\kms{km~s$^{-1}$}

\begin{document}

   \title{A possible relation between global CO excitation and massive molecular outflows in local ULIRGs}

   \subtitle{}

   \author{I. Montoya Arroyave
                \inst{1}
        \and C. Cicone\inst{1}
        \and P. Andreani\inst{2}
        \and A. Weiss \inst{3}
        \and C. De Breuck\inst{2}
        \and A. Lundgren\inst{4,2}
        \and P. Severgnini\inst{5}
        \and B. Hagedorn\inst{1}
        \and Rubinur K. \inst{1}
        \and B. Baumschlager\inst{1}
        \and E. Makroleivaditi\inst{3,6}
          }
   \institute{Institute of Theoretical Astrophysics, University of Oslo,
   P.O. Box 1029, Blindern, 0315 Oslo, Norway \\  
              \email{isabemo@uio.no}
          \and European Southern Observatory, Karl-Schwarzschild-Strasse 2, 85748 Garching, Germany 
          \and Max-Planck-Institut fur Radioastronomie, Auf dem Hugel 69, D-53121 Bonn, Germany 
          \and Aix Marseille Université, CNRS, LAM, Marseille, F-13388, France 
          \and INAF - Osservatorio Astronomico di Brera, Via Brera 28, I-20121 Milano, Italy 
          \and Rheinische Friedrich-Wilhelms-Universitat Bonn, Regina-Pacis-Weg 3, 53113 Bonn, Germany 
        }

   \date{Received 2023-10-17; Accepted 2024-02-22}

 
  \abstract
   {
   Local ultra-luminous infrared galaxies (ULIRGs) have been observed to host  ubiquitous molecular outflows, including the most massive and powerful ever detected. These sources have also exceptionally excited global, galaxy-integrated CO ladders.
   A connection between outflows and molecular gas excitation has however never been established, since previous multi-$J$ CO surveys were limited in spectral resolution and sensitivity and so could only probe the global molecular gas conditions.
   In this work, we address this question using new, ground-based, sensitive heterodyne spectroscopy of multiple
   CO rotational lines (up to CO(7--6)) in a sample of 17 local ULIRGs. We used the Atacama Pathfinder Experiment (APEX) telescope to survey the CO($J_{\rm up}\geq4$) lines at a high signal-to-noise ratio, and complemented these data with CO($J_{\rm up}\leq3$) APEX and Atacama Large Millimeter Array (ALMA and ACA) observations presented in \citet{MontoyaArroyave+23}.  
   We detected a total of 74 (out of 75) CO lines, with up to six transitions per source.
   The resulting CO spectral line energy distributions (SLEDs) show a wide range in gas excitation, in agreement with previous studies on ULIRGs. Some CO SLEDs  peak at $J_{\rm up }\sim3,4$, which we classify as `lower excitation', while others plateau or keep increasing up to the highest-$J$ CO transition probed, and we classify these as `higher excitation'. 
   Our analysis includes for completeness the results of CO SLED fits performed with a single large velocity gradient component, but our main focus is the investigation of possible links between global CO excitation and the presence of broad and/or high-velocity CO spectral components that can contain outflowing gas.
   We discovered an increasing trend of line width as a function of $J_{\rm up}$ of the CO transition, which is significant at the $4\sigma$ level and appears to be driven by the {eight} sources that we classified as `higher excitation'.
   We further analysed such higher-excitation ULIRGs, by performing a decomposition of their CO spectral profiles into multiple components,  {and derived CO ladders that are} clearly more excited for the spectral components characterised by higher velocities and/or velocity dispersion. {Because these sources are known to host widespread molecular outflows, we favour an interpretation whereby the highly excited CO-emitting gas in ULIRGs resides in galactic-scale massive molecular outflows whose emission fills a large fraction of the beam of our APEX high-$J$ CO observations.} 
   {On the other hands, our} results challenge {alternative} scenarios for which the high CO excitation in ULIRGs can be explained by classical component of the ISM, such as {photon- or X-ray dominated regions around the nuclear sources}. 
   }


   \maketitle
%

\section{Introduction}\label{sec:intro}
In the local universe ($z \leq 0.2$), most (ultra) luminous infrared galaxies ((U)LIRGs), defined by their high infrared luminosities ($L_{\rm IR}(8-1000{\rm \mu m}) \geq 10^{11}{\rm L}_{\odot}$), correspond to galaxy mergers that can trigger accelerated star formation (starbursts, SBs) or accretion of matter onto the supermassive black hole (SMBH) in the center of the galaxy (active galactic nuclei, AGN) \citep{Sanders+Mirabel96, Genzel+98, Lonsdale+06, Perez-Torres+21, U22}.
The stellar and AGN feedback in these sources generate the most powerful galactic outflows known, which can embed gas in different phases and affect the host galaxy from scales of a few parsecs in the interstellar medium (ISM) to several tens of kiloparsecs out to the circumgalactic medium (CGM) \citep{Cicone+15}.

The molecular gas phase of the ISM is the raw material from which stars form, and so constraining its physical properties in different gas environments, and in particular diffuse and extended outflowing gas, is important to understand the evolution of galaxies.
Being rich in dense gas and dust, (U)LIRGs are ideal targets for (sub)millimeter observations of the molecular gas phase.
Indeed, in the past decade, thanks to {\it Herschel} and ground-based sub-mm observatories, massive molecular (H$_2$) outflows have been detected in the majority of local (U)LIRGs (see \cite{Veilleux+20} for a review).
In the recent years, observations of different gas tracers, such as CO, HCN, HCO+ (among others) of these sources have also allowed to study the implications of galactic outflows in the galaxy evolution scheme.
Galactic outflows  may be the culprits of either removing large fractions of gas from the host galaxy, and hence decrease the star formation rate (SFR; negative feedback; e.g., \citealt{DiMatteo+05,Tumlinson+17}), or compressing the available gas, resulting in increased SFR (positive feedback; e.g., \citealt{Maiolino+17,Gallagher+19}).
Despite extensive theoretical and computational work aimed at understanding galaxy-scale outflows, the propagation of cold outflows to scales of $\sim10$~kpc or larger, {and their interaction with the gas phases embedded in the} circumgalactic medium (CGM), is particularly difficult to model because of high numerical resolution requirements in cosmological simulations \citep{Hummels+19,Suresh+19,VandeVoort+19,Schimek+23}.
Therefore, further observational constraints of cold gas outflows, including their most extended components {that are often faint or missed in} high-resolution interferometric observations, are strongly needed to 
inform theoretical models.


Given that near-IR and mid-IR H$_2$ lines are not sensitive to the cold (i.e., $T<100~$K) molecular phase, we need proxy tracers.
The most common molecular tracer is the CO molecule, as it is produced in similar conditions to molecular hydrogen and is collisionally thermalised by H$_2$ even at low densities \citep{Bothwell+13}.
Its low-$J$ rotational transitions are easily excited as they have low critical densities ($\sim300$ cm$^{-3}$ for CO(1--0)), thus sensitive to the total gas reservoir.
The higher-$J$ transitions have larger critical densities ($\sim10^4$ cm$^{-3}$ for CO(5--4)), and arise from denser and/or warmer regions in the gas.
As such, observations of the different rotational transitions of a molecule can be used as a probe of the physical conditions of the gas, as its excitation depends on a competition between radiative and collisional transitions, with the latter depending on the temperature and density of the gas.
The shape of the rotational ladder of CO (also referred to as the spectral line energy distribution, SLED), {which is} a plot of the CO($J_{\rm up}\rightarrow J_{\rm up}-1$) integrated line fluxes as a function of the upper rotational quantum number $J_{\rm up}$, is therefore sensitive (albeit in a degenerate way) to the H$_2$ gas density and temperature.

It has been proposed that the ISM in ULIRGs (as well as other galaxy populations) is composed of a low-excitation cold component phase that traces most of the H$_2$ gas mass, embedded in a warm, turbulent, and diffuse component, traced by higher excitation CO emission lines \citep{Imanishi+23,Aalto+95}.
In this sense, the CO SLEDs may be broken down into three different regions \citep[e.g.,][]{Rosenberg+15, Decarli+20,Boogaard+20,Esposito+22}: (i) the low-$J$ transitions ($J_{\rm up}\lesssim3$) that trace the bulk of the molecular gas mass with cold temperatures and low densities, (ii) the mid-$J$ transitions ($4\lesssim J_{\rm up}\lesssim9$) which typically trace warmer and denser gas related to star formation, the photon-dominated regions (PDRs), and (iii) the high-$J$ transitions ($J_{\rm up}\gtrsim10$) tracing extreme conditions such as shocks, X-ray dominated regions (XDRs) {or cosmic ray-dominated regions (CRDRs)} often linked to AGN activity.
Combining information from the different rotational transitions of the CO molecule leads to a better understanding of the heating mechanisms behind the observed excitation of the ISM in galaxies.

Because SBs and AGNs introduce intense mechanical and radiative feedback into the ISM, it is expected that these have strong effects on the gas excitation, and thereby, on CO SLEDs. The incidence of massive molecular outflows in ULIRGs is extremely high, $>$70\% \citep{Veilleux+13, Lamperti+22}, pointing to extremely powerful feedback mechanisms at work in these sources.
As a matter of fact, studies performed on local (U)LIRGs over the last decade thanks to \textit{Herschel}, have shown that the global CO SLEDs of these sources peak at $J\geq 6$, and may even remain flat up to $J\simeq10$ \citep{Kamenetzky+14,Lu+14,Rosenberg+15,Greve+14,Liu+15,Mashian+15}, while the CO ladder of normal disc galaxies have been observed to peak at lower $J\simeq3$ \citep{Dannerbauer+09,Crosthwaite-Turner+07}.

{Given the high incidence ($\gtrsim70$~\%) of massive outflows in ULIRGs, their} extreme CO SLEDs may be connected to the powerful feedback mechanisms at work in these galaxies. However, a clear connection between molecular outflows and CO gas excitation has never been established: on the one hand, CO excitation studies of outflows are sparse and only sample a few, low-$J$ CO lines (see references and discussion in \cite{MontoyaArroyave+23}; on the other hand, the {\it Herschel} spectroscopic observations of multiple CO transitions, which have been used to probe the CO SLEDs of ULIRGs up to very high $J_{\rm up}$, are not adequate to study the connection between outflows and CO excitation due to their low spectral resolution, 
which prevent a reliable study of the CO line profile.

It is worth noting, however, the diversity in the observed CO SLEDs for a given galaxy population \citep{Narayanan+14}. 
Indeed, while the CO fluxes measured in ULIRGs are characterized by higher excitation than those measured in normal SF disc galaxies, and trends have been found, for instance, between excitation and high IR luminosity and/or high dust temperature (traced by the $60\mu{\rm m}/100\mu{\rm m}$ color index) \citep[see e.g.][]{Lu+14,Rosenberg+15,Pearson+16}, their SLEDs span a wide range of excitation for $J_{\rm up}\gtrsim5$ lines \citep[see e.g.,][]{Papadopoulos+12a,Mashian+15}.
Similarly, observed CO SLEDs of dusty star forming galaxies selected through their sub-millimeter fluxes (also known as sub-millimeter galaxies, SMGs), which are commonly considered ULIRG-type galaxies at high-$z$, also show a wide variety in their measured CO excitation \citep[see e.g.,][]{Harrington+21,Yang+17,Spilker+14,Bothwell+13,Carilli+Walter13}.
Such galaxy selections based on arbitrary luminosity- or color-cuts can result in heterogeneous physical conditions. Consequently, no `template' SLED exists for a given population such as (U)LIRGs, and we must pay particular attention when deriving average gas conditions for galaxy samples. 

While modeling individual CO SLEDs is necessary for understanding the different physical conditions that galaxies host, we can still, however, study statistical trends in galaxy populations.
ULIRGs are known to host powerful galactic outflows, leading to large velocity gradients (LVG) which may be related to the high excitation CO SLEDs observed in these sources. 
Although spatially resolved observations are needed to determine and characterize outflowing and non-outflowing material, single-pointing single-dish observations that deliver the total integrated molecular line emission constitute an essential first step for recovering the total flux of the poorly explored high-$J$ CO lines, whose luminosity is unknown especially for the broad line wings that can trace outflows. This was the aim of our APEX observing campaign targeting 40 local (U)LIRGs, complemented by archival APEX and ALMA data, whose first results based on the low-$J$ ($J\leq3$) CO and [CI](1-0) lines have been reported in \cite{MontoyaArroyave+23}. 

One of the results presented in \cite{MontoyaArroyave+23} was the measurement of extremely high galaxy-integrated low-$J$ CO ratios in local (U)LIRGs, significantly offset from those measured in local massive main-sequence (MS) galaxies.
However, from a spectral decomposition of the CO lines into low-velocity/narrow and high-velocity/{broad} line components, which, in first approximation, are dominated respectively by disc and outflowing gas, we found no evidence of consistently higher CO excitation in the latter, based on CO transitions up to $J_{\rm up}=3$.
These results highlighted the need for higher-$J$ CO lines to better constrain the excitation in these sources, as CO($J_{\rm up}<3$) lines have little diagnostic value concerning the molecular ISM excitation in local (U)LIRGs, contrary to what has been found for normal star-forming galaxies in the local Universe \citep[e.g.,][]{Leroy+22,Lamperti+20}.

{In this follow-up paper, we} extend the analysis performed in \cite{MontoyaArroyave+23} for a subsample of 17 local ULIRGs, with high-sensitivity single-pointing APEX observations of CO lines up to $J_{\rm up}=7$. This effort is the first of its kind as it can rely on sensitive, high frequency (up to $\sim710$~GHz), sub-millimetric ground-based single-dish data for a conspicuous sample of local ULIRGs, obtained with wide bandwidths and high spectral resolution, reduced and analysed in a homogenous way, by paying particular attention to aperture effects and where the few scans with poor baselines have been removed from the spectra.
The paper is organized as follows.
In Sect.~\ref{sec:observations} we describe the sample, the observing strategy, and the data reduction and analysis. 
In Sect.~\ref{sec:excitation} we construct the CO SLEDs and investigate the ISM excitation and physical properties, where they are also contextualized through a comparison with relevant results from the literature.
In the same section we use radiative transfer models to extract the average density, temperature and column density of the global ISM observed in our sample, and comment on caveats of deriving global gas properties.
In Section~\ref{sec:link_exc_out} we explore {correlations} between the higher CO excitation of ULIRGs and {the appearance of broad and/or high-velocity spectral components, to investigate the links with} massive molecular outflows.
Finally, we summarize our results and conclusions in Sect.~\ref{sec:conclusions}.
Throughout this work, we adopt a $\Lambda$-cold dark matter cosmology, with $H_0=67.8$ \kms Mpc$^{-1}$, $\Omega_{\textrm{M}}=0.307,$ and $\Omega_{\Lambda}=0.693$ \citep{Planck_cosmology+14}.




\section{Observations and Data}\label{sec:observations}

\subsection{The sample}\label{sec:sample}

\begin{table*}[tbp]
        \centering \small
        \caption{Galaxies analyzed in this work along with some general properties.}\label{table:source_list}
        \begin{tabular}{lccccccccc}
                \hline
                \hline
                Galaxy name           &  z        &  $D_{L}$  &     R.A.      &      Dec        & $\alpha_{\textrm{AGN}}$  & $\log L_{\textrm{IR}}$ & $\log L_{\textrm{AGN}}$ & SFR &  Ref.  \\
                                      &           &   [Mpc]   &               &                 &             & [$L_{\odot}$] & [$L_{\odot}$]           & [M$_{\odot}\textrm{yr}^{-1}$]  &   \\
                (1)                   &  (2)      &  (3)      &    (4)        &     (5)         & (6)         &  (7)    & (8)           & (9)           & (10)                               \\
                \hline
                IRAS F01572+0009      & 0.1631    & 806.66    & 01:59:50.253  & +00:23:40.87    &  0.65       & 12.62   & 12.49         & 150$\pm$60    &  $\beta$    \\
                IRAS 03521+0028       & 0.1519    & 746.13    & 03:54:42.219  & +00:37:03.41    &  0.06       & 12.52   & 11.39         & 309$\pm$120   &  $\gamma$   \\
                IRAS F05189$-$2524    & 0.0426    & 194.52    & 05:21:01.392  & $-$25:21:45.36  &  0.72       & 12.16   & 12.07         & 40$\pm$16     &  $\beta$    \\
                IRAS 06035$-$7102     & 0.0795    & 372.55    & 06:02:54.066  & $-$71:03:10.48  &  0.60       & 12.22   & 12.06         & 70$\pm$30     &  $\gamma$   \\
                IRAS 06206$-$6315     & 0.0924    & 436.81    & 06:21:01.210  & $-$63:17:23.5   &  0.43       & 12.23   & 11.92         & 100$\pm$40    &  $\gamma$   \\
                IRAS 08311$-$2459     & 0.1005    & 477.67    & 08:33:20.600  & $-$25:09:33.7   &  0.79       & 12.50   & 12.46         & 70$\pm$30     &  $\gamma$   \\
                IRAS 09022$-$3615     & 0.0596    & 275.47    & 09:04:12.689  & $-$36:27:00.76  &  0.55       & 12.29   & 12.09         & 90$\pm$30     &  $\beta$    \\
                IRAS 10378+1109       & 0.1363    & 663.02    & 10:40:29.169  & +10:53:18.29    &  0.30       & 12.31   & 11.85         & 140$\pm$60    &  $\gamma$   \\
                IRAS F12112+0305      & 0.07309   & 341.01    & 12:13:45.978  & +02:48:40.4     &  0.18       & 12.32   & 11.63         & 170$\pm$70    &  $\beta$    \\
                IRAS F13451+1232      & 0.1217    & 586.51    & 13:47:33.425  & +12:17:24.32    &  0.81       & 12.32   & 12.29         & 41$\pm$15     &  $\beta$    \\
                IRAS 16090$-$0139     & 0.1336    & 648.77    & 16:11:40.432  & $-$01:47:06.56  &  0.43       & 12.55   & 12.25         & 200$\pm$80    &  $\gamma$   \\
                IRAS 17208$-$0014     & 0.0428    & 195.47    & 17:23:21.920  & $-$00:17:00.7   & $\leq$0.05  & 12.39   & 11.15         & 230$\pm$90    &  $\beta$    \\
                IRAS 19254$-$7245     & 0.06149   & 284.58    & 19:31:21.400  & $-$72:39:18.0   &  0.74       & 12.09   & 12.02         & 32$\pm$12     &  $\gamma$   \\
                IRAS F20551$-$4250    & 0.0430    & 196.41    & 20:58:26.781  & $-$42:39:00.20  &  0.57       & 12.05   & 11.87         & 48$\pm$19     &  $\beta$    \\
                IRAS F22491$-$1808    & 0.0778    & 364.16    & 22:51:49.264  & $-$17:52:23.46  &  0.14       & 12.84   & 12.05         & 590$\pm$220   &  $\beta$    \\
                IRAS F23060+0505      & 0.1730    & 860.75    & 23:08:33.952  & +05:21:29.76    &  0.78       & 12.53   & $\cdot\cdot\cdot$ & 80$\pm$30 &  $\delta$   \\
                IRAS F23128$-$5919    & 0.0446    & 203.95    & 23:15:46.749  & $-$59:03:15.55  &  0.63       & 12.03   & 11.89         & 40$\pm$15     &  $\beta$    \\
                \hline
        \end{tabular}

        \tablefoot{(1) Source name. (2) Redshift. (3) Luminosity distance. (4) Right ascension. (5) Declination. (6) Reported fractional contribution of the AGN to the bolometric luminosity in the reference papers ($\alpha_{\textrm{AGN}} = L_{\rm AGN} /L_{\rm bol} $). (7) Infrared luminosity $(8-1000$  $\mu \textrm{m})$. (8) AGN luminosity. (9) Star formation rate.\\
        \textbf{References.} $\beta$: \citet{Veilleux+13}, $\gamma$: \citet{Spoon+13}, {$\delta$}: \citet{Fluetsch+19}.}

\end{table*}

The parent sample of our study, i.e. the 40 (U)LIRGs presented in \cite{MontoyaArroyave+23}, is representative the local ULIRG population in terms of physical properties, e.g., star formation rate (SFR), AGN luminosity ($L_{\rm AGN}$), and AGN fractions ($\alpha_{\textrm{AGN}} \equiv L_{\rm AGN} /L_{\rm bol}$).
We now extend the analysis to higher-$J$ CO data ($4\leq J_{\rm up} \leq 7$) for 17 ULIRGs, which span the entire range of $L_{\rm AGN}$ and SFR of the parent sample, in the redshift range $0.04<z<0.18$.
{Previous {\it Herschel} OH119$\mu$m observations, which are discussed in more detail in \cite{MontoyaArroyave+23}, have reliably detected molecular outflow signatures in 11 out of the 17 galaxies studied in this paper (65\% detection rate, which is typical for ULIRGs).}
Table \ref{table:source_list} lists the sub-sample analyzed throughout this paper.


\subsection{Observing strategy}\label{sec:obs_strategy}
\subsubsection{Low-{\it J} CO data}
{A full description of low-$J$ ($J_{\rm up}\leq3$) CO data} was given in \cite{MontoyaArroyave+23}, {hence here we only summarise the main points of the data acquisition and processing.}
The low-$J$ CO data combine proprietary and archival single-dish (APEX) and archival interferometric (ACA, ALMA and IRAM PdBI) observations.
Since we aim to include in our analysis the most extended and diffuse molecular gas components of the ISM of these sources, we gave priority to observations with the highest sensitivity to large-scale structures. In this sense, when data from multiple instruments are available for the same source and transition, we set the following priorities: (1) our own APEX proprietary (PI) data, (2) archival APEX or IRAM PdBI data, (3) archival ACA data, and, lastly, (4) archival ALMA data. This means that, when available, we use our own APEX PI observations since they were designed expressly for our science goals, although this is not always possible (for example, APEX does not have a 100~GHz receiver providing CO(1--0) line coverage).
For the sub-sample of 17 ULIRGs presented in this work, we have the following low-$J$ CO data coverage: 
CO(1--0) spectral line data for 10 galaxies, CO(2--1) line observations for all 17 sources, and lastly, 16 galaxies with CO(3--2) line coverage.

The spectral setup and sensitivity goals of the APEX PI data are described in detail in \cite{MontoyaArroyave+23}.
The receivers used were SEPIA180, SEPIA3445 and nFLASH230 depending on the observed frequency of interest, and all data were reduced and analyzed in a consistent way.
For the APEX archival data, a compendium of different project datasets were used, using mainly the SHeFI and nFLASH receivers.
For the single-dish APEX data, the beam sizes are $\sim27$ arcsec for the CO(2--1) line, and $\sim20$ arcsec for the CO(3--2) line. The apertures used for extracting the spectra for the interferometric data are reported in \cite{MontoyaArroyave+23}.
For all single-dish data, PI and archival, we adopted the same reduction and analysis steps, similar to the one explained in Sect. \ref{sec:data_reduction} for the high-$J$ CO data presented in this work.

The archival ALMA and ACA data were gathered from different projects with different array configurations, leading to different angular resolutions and maximum recoverable scales.
We re-reduced all these data paying particular attention to recovering the extended and diffuse emission (which may be affected by the choice of the cleaning mask, for example), and extracted the final spectrum from an aperture that was size-matched to maximize the recovered flux \citep[more details can be found in][]{MontoyaArroyave+23}.
For the sources where only high-resolution ALMA observations were available, we applied {\it uv}-tapering to enhance the sensitivity to extended structures.
In the subsample analyzed in this work, only the CO(1--0) line observation of the source IRAS F01572+0009 
{is a high resolution ALMA dataset that may be biased against the most extended and diffuse components}.
We are aware, however, that {\it uv}-tapering does not overcome the issue of missing flux from faint extended structures due to the poor sampling of short {\it uv} baselines inherent to interferometric observations.

\subsubsection{Higher-{\it J} CO data}

All mid-$J$ CO rotational transitions (with $J_{\rm up}=4,5,6,7$) were observed with the APEX telescope between October 2020 and June 2021 (project ID \texttt{E-0106.B-0674}, PI: I. Montoya Arroyave) for the sources presented in Table \ref{table:source_list}.
Our observing strategy was to reach a line peak-to-rms ratio of S/N $\sim 5$ on the expected peak flux density in velocity channels $\delta v \sim 50-100$ \kms, assuming a line width equal to that measured in the CO(2--1) line observation for each source.

The CO(4--3) ($\nu^{\rm rest}_{\mathrm{CO(4-3)}}$ = 461.041 GHz) and CO(5--4) ($\nu^{\rm rest}_{\mathrm{CO(5-4)}}$ = 576.268 GHz) observations were carried out with the nFLASH460 receiver, and the CO(6--5) ($\nu^{\rm rest}_{\mathrm{CO(6-5)}}$ = 691.473 GHz) and CO(7--6) ($\nu^{\rm rest}_{\mathrm{CO(7-6)}}$ = 806.652 GHz) observations were performed with the SEPIA660 receiver.
Both instruments are frontend heterodyne with dual-polarization sideband-separating (2SB) receivers, and each sideband (upper -USB-, and lower -LSB-) is covered by a fast Fourier transform spectrometer (FFTS) unit.
The nFLASH460 receiver covers a frequency range between 378 and 508 GHz, and has a IF coverage between 4 and 8 GHz, with an instantaneous coverage of 4 GHz per sideband and a 12 GHz gap between the sidebands.
The SEPIA660 receiver can be tuned in the frequency range $597-725$~GHz, and it has two IF outputs per polarization, each covering $4-12$~GHz separated by a 16~GHz gap, leading to a total of up to $\Delta \nu = 32$~GHz IF bandwidth \citep{Meledin+22}. 
The average noise temperatures of the receivers are ($T_{\textrm{rx}}$) of $\sim 150$~K for nFLASH460, and between $\sim 75-125$ K for SEPIA660 \citep{Belitsky+18}.
The beam sizes for the mid-$J$ CO observations are $\sim14,\,11,\,9$ and $8$ arcsec for the CO(4--3), CO(5--4), CO(6--5) and CO(7--6) lines, respectively.

For our observations, each sideband spectral window was divided into 4096 channels ($\sim 977$ kHz per channel), corresponding to a resolution in velocity units at the range of redshifts covered by our sample of $\sim0.6$ \kms~for the nFLASH460 obervations, and $\sim0.4$ \kms~for the SEPIA660 observations.
The telescope was tuned to the expected observed frequency\footnote{For most observations carried out in 2021 the telescope was tuned to the expected observed frequency plus a 2 GHz shift. This allowed us to probe more of the baseline continuum around the emission line to perform a better continuum baseline fit and subtraction.} for each transition, with wobbler-switching symmetric mode with $100''$ chopping amplitude and a chopping rate of $R=0.5 \textrm{ Hz}$. 
The data were calibrated using standard procedures (see Section \ref{sec:data_reduction}). 
The on-source integration times (without overheads) varied from source to source, from $\sim5$ minutes up to $\sim2$ hours.
During the observing runs, the PWV varied from $0.2 < {\rm PWV}\textrm{ [mm]} < 1.0$. 

Our final data coverage for the mid-$J$ CO lines is as follows: CO(4--3) line observations for 16 sources, among which there is a non-detection due to a nearby atmospheric line (IRAS~06035-7102); CO(5--4) line detections in two sources, CO(6--5) line detections in 11 sources, and lastly, three sources with CO(7--6) line measurements.
Therefore, in total, we detected 74 CO lines with $J_{\rm up} \leq7$ out of the 75 observed lines. Our data set includes between three and six CO lines per source. 
We adopted the same reduction and analysis steps for all data, which are summarized below in Sect.~\ref{sec:data_reduction}.


\subsection{Data reduction}\label{sec:data_reduction}
The low-$J$ CO lines data reduction is explained in detail in \cite{MontoyaArroyave+23}, both for the proprietary APEX data, and the archival ALMA/ACA data.
For the high-$J$ CO APEX data, we follow a similar procedure as that of the low-$J$ CO APEX data.
Here, we briefly go through the steps of such data reduction process.

We reduced the data using the \texttt{GILDAS/CLASS} software package\footnote{https://www.iram.fr/IRAMFR/GILDAS/}.
We began by isolating all the spectral scans for each source and transition and inspected them individually in each polarization, we then discarded those scans that were affected by baseline instabilities and other instrumental issues.
We subtracted the baselines in individual scans after masking the spectral range corresponding to $v \in (-500, 500)$ \kms, in order to avoid the expected line emission, and followed by co-adding all scans and then subtracting a linear baseline to the averaged spectrum.
We fitted a single Gaussian function to the line profile, and based on the resulting fit parameters (line width and central velocity), we adjusted the central mask to the velocity channels of the detected line emission.
Lastly, we produced the final spectra and exported them with a common spectral resolution of $\delta v \approx 5$ \kms, in order to allow for further rebinning in the spectral fitting phase. The spectra were imported into Python\footnote{https://www.python.org}, where the data analysis was performed.

Since the output signal from the APEX telescope corresponds to the antenna temperature corrected for atmospheric losses ($T_{A}^{*}$) in units of Kelvin [K], it must be multiplied by a calibration factor (or telescope efficiency) in order to obtain the flux density in units of Jansky [Jy]\footnote{Strictly speaking, the units are Jy/beam as we are not integrating spatially. Nonetheless, we are assuming that all the emission is contained in the beam, hence the total flux density is in Jy}.
For the nFLASH460 receiver, the average Kelvin to Jansky conversion factor measured during our observation period is $63 \pm 11$ $\mathrm{Jy~K^{-1}}$, and for SEPIA660 it is $57 \pm 5$ $\mathrm{Jy~K^{-1}}$.
{The APEX observations that are presented for the first time in this work, covering the mid-$J$ CO lines ($4\leq J_{\rm up}\leq 7$) in 17 sources, are shown in Figures~B1 to B17 in the Appendix.}

\subsection{Data analysis}\label{sec:data_analysis}
The analysis performed in this work is based on galaxy-integrated molecular line spectra, and as such, it is aimed at estimating the average physical conditions of the molecular ISM in our sample of ULIRGs using source-averaged CO SLEDs.

Once exported in Python, all CO line spectra were fit using \texttt{mpfit} \footnote{Open-source algorithm adapted from the IDL version; see \href{https://github.com/segasai/astrolibpy/blob/master/mpfit/}{https://github.com/segasai/astrolibpy/blob/master/mpfit/}}.
This code uses the Levenberg-Marquardt technique to solve the least-squares problem in order to fit data (the observed spectrum) with a model (a multi-Gaussian function).
Each Gaussian component is characterized by an amplitude, a peak position ($v_{\rm cen}$), and a velocity dispersion ($\sigma_v$, or, equivalently, the full width at half maximum, FWHM). 
We allowed the fit to use up to a maximum of three Gaussian functions to reproduce the observed global line profiles, so that line asymmetries and broad wings are properly captured with separate spectral components when the S/N is high enough (see e.g., source IRAS 09022-3615, Figure \ref{fig:spec09022}).
At this stage, we do not assign any physical meaning to the different Gaussian components fitted to each transition, nor force the fit to use the same Gaussians for all CO transitions of the same source, 
as our goal is to simply reproduce the total spectral line profiles.



We verified that the velocity-integrated line fluxes measured through the Gaussian fits are consistent with those calculated by directly integrating the spectra within $v \in (-1000, 1000)$ \kms, after setting a {signal-to-noise threshold} of S/N$>1.5$ per channel. We find that the two methods give consistent results within the statistical errors. 
The total line fluxes are reported in Table \ref{table:fluxes}. 

We then converted the {velocity-integrated} fluxes to line luminosities following \cite{Solomon+97}:
\begin{equation}
        L{'}_{\rm line}\,\mathrm{ [K\,km\,s^{-1}\,pc^{2}]} = 3.25 \times 10^7\frac{D^{2}_{L}}{\nu^{2}_{\mathrm{obs}}(1+z)^3}\, \int{S_{v}\, {{\rm d}v}}\, ,
\end{equation}
where $D_{L}$ is the luminosity distance measured in Mpc, $\nu_{\textrm{obs}}$ is the corresponding observed frequency in GHz, and $\int S_{v}\,\mathrm{d}v$ is the total integrated line flux in Jy~\kms. 
In Table \ref{table:luminosities} we report the luminosities calculated for the different transitions.

From the line luminosities we can compute the CO line ratios with respect to the CO(1--0) line luminosity, {where we use the following notation}:
\begin{equation}
    r_{J1}\equiv L'_{{\rm CO}(J\rightarrow J-1)}/L'_{\rm CO(1-0)}
\end{equation}
where $J$ corresponds to the upper rotational quantum number, $J_{\rm up}$.
The average global ratios calculated for our sample are reported in Table \ref{table:avg_ratios}.

\begin{table}[tbp]
\small
        \centering 
        \caption{Average CO line ratios of the ULIRGs in our sample.}\label{table:avg_ratios}
        \begin{tabular}{cccc}
                \hline
                \hline
                Ratio $^{a}$       &     Mean     & Error $^{b}$ &   $N^{c}$   \\
                \hline
                $r_{21}$           &    $1.10$    &    $0.03$    &    10       \\ 
                $r_{31}$           &    $0.77$    &    $0.02$    &    10       \\ 
                $r_{41}$           &    $0.54$    &    $0.03$    &    9        \\ 
                $r_{51}$           &   ($0.42$)   &    $-$       &    2        \\ 
                $r_{61}$           &    $0.17$    &    $0.007$   &    6        \\ 
                $r_{71}$           &   ($0.58$)   &    $-$       &    2        \\ 
                \hline
        \end{tabular}

        \tablefoot{$^{(a)}$CO luminosity ratios, $r_{J1}\equiv L'_{{\rm CO}(J\rightarrow J-1)}/L'_{\rm CO(1-0)}$. 
        $^{(b)}$The error is the uncertainty of individual measurements divided by the square root of the number of data points in that sample. $^{(c)}$Total number of measurements available for each luminosity ratio. 
        The $r_{21}$ and $r_{31}$ ratios were studied in \cite{MontoyaArroyave+23} for the entire parent sample of 40 local (U)LIRGs, with average values $\langle r_{21}\rangle=1.12\pm0.03$ and $\langle r_{31}\rangle=0.81\pm0.02$, well in agreement with those measured in this work.
        {The average} ratios $r_{51}$ and $r_{71}$ are computed solely from two measurements, therefore they {are affected by a large variance}.
        }

\end{table}

\section{CO excitation analysis}\label{sec:excitation}

\subsection{CO SLEDs}\label{sec:sleds}
\begin{figure*}[tbp]
        \centering
        \includegraphics[width=.99\textwidth]{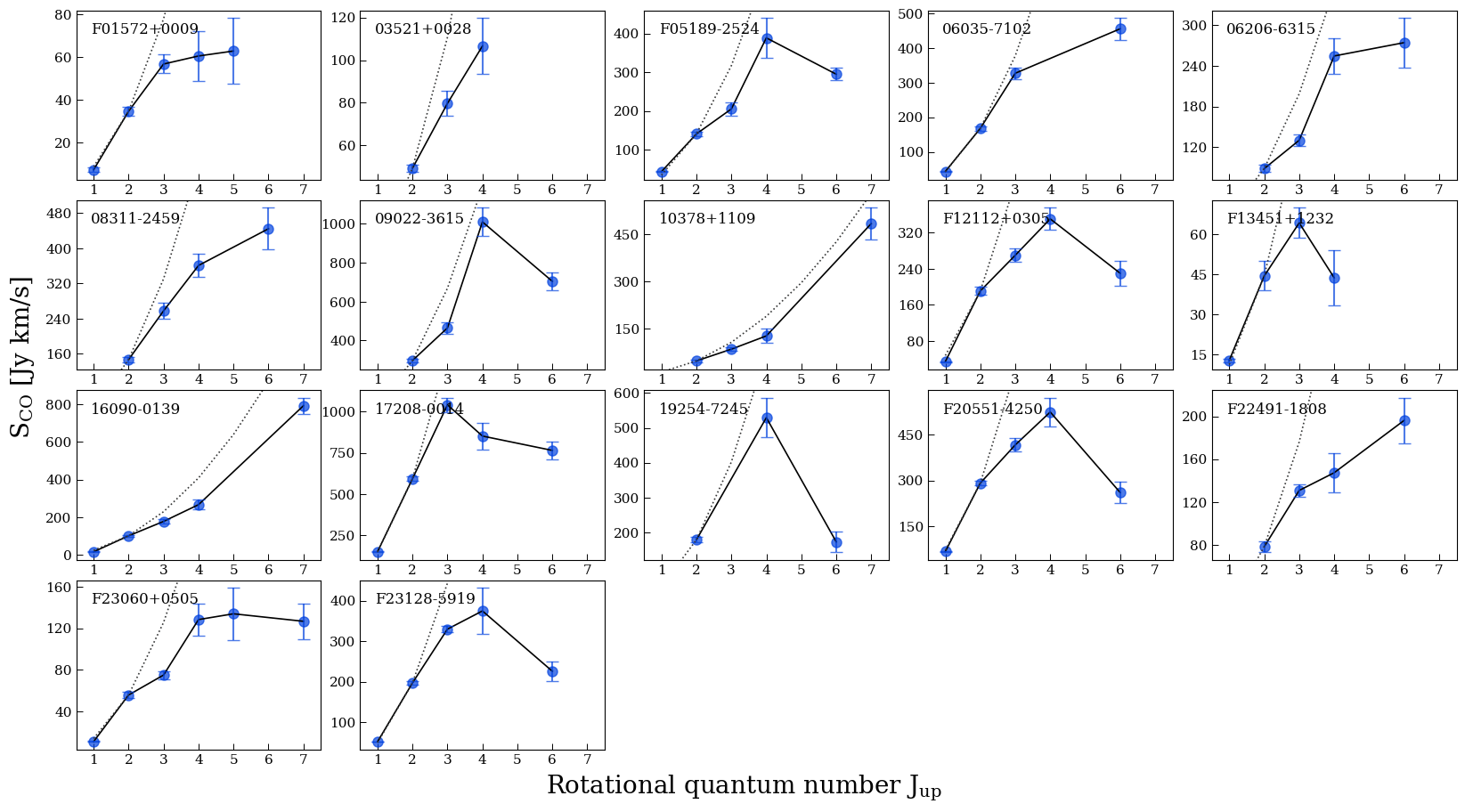}
        \caption{CO spectral line energy distributions of the ULIRGs in our sample. The blue circles with the error bars are the velocity-integrated line fluxes in units of Jy~\kms. The dotted gray lines show the thermalized, {optically thick limit (in all lines)}, SLED profiles, normalized to the CO(2--1) transition {(which is the lowest-$J$ CO transition that is available for all sources)}. The source IDs are reported on the top-left corner of each panel.}
        \label{fig:global_CO_SLEDs_grid}
\end{figure*}

Our results are displayed in Figure~\ref{fig:global_CO_SLEDs_grid} in the form of observed CO SLEDs (using integrated flux density in Jy km/s vs $J_{\rm up}$), for all 17 ULIRGs. This plot demonstrates the wide range of gas excitations in the ULIRG population, visible from the different shapes of the observed CO ladders.
Already from a visual inspection, we can observe sources with SLEDs peaking at $J_{\rm up}\sim 3,4$ (e.g., IRAS F05189-2524, IRAS 09022-3615, IRAS F12112+0305, IRAS F13451+1232, IRAS 17208-0014, IRAS 19254-7245, IRAS F20551-4250 and IRAS F23128-5919), others showing a plateau-like feature from $J_{\rm up} \gtrsim 3$ (IRAS F01572+0009 and IRAS F23060+0505), and lastly, sources for which no plateau nor turn-over point is observed up to the highest detected CO transition (IRAS 03521+0028, IRAS 06035-7102, IRAS 06206-6315, IRAS 08311-2459, IRAS 10378+1109, IRAS 16090-0139 and IRAS F22491-1808).
The peak of the CO SLED sheds light on the gas excitation conditions, since it directly depends on the gas kinetic temperature and density.
For source IRAS~03521+0028, with only three detected CO ($J_{\rm up} \leq 4$) lines and no visible turn-over, we cannot constrain the H$_2$ gas properties.

We also show, indicated by a gray dotted curve in each panel of Figure~\ref{fig:global_CO_SLEDs_grid}, the trend expected for thermalized CO emission with fluxes rising as the square of $J_{\rm up}$ (assuming all lines are in the Rayleigh-Jeans regime).
Star forming ULIRGs are expected to have a larger fraction of dense, excited {CO} gas that may be thermalized up to $J_{\rm up} = 3$ or beyond \citep[e.g.,][]{Weiss+07, Papadopoulos+12a}, in contrast to the low-density, low-excitation gas dominating the CO emission from more quiescent galaxies \citep{Braine+93,Crosthwaite-Turner+07,Boogaard+20}.
Indeed, Figure~\ref{fig:global_CO_SLEDs_grid} shows that CO(3--2) is only weakly sub-thermally excited for most sources in our sample. The average $r_{31}$ line ratio is $0.77\pm0.02$ (see Table~\ref{table:avg_ratios}), consistent with moderately sub-thermally excited lines.
Conversely, CO($J_{\rm up}\geq4$) lines are significantly sub-thermally excited in most sources. 
For the higher-$J$ transitions, we derived average galaxy-integrated line ratios of $r_{41}=0.54\pm0.03$, $r_{51}=0.42\pm0.07$, $r_{61}=0.17\pm0.007$ and $r_{71}=0.58\pm0.04$.
It is worth noting, however, the lower statistics for the $J_{\rm up}\geq5$ ratios, and in particular $J_{\rm up}=5,7$, for which we only have two measurements each; as such, the mean $r_{51}$ and $r_{71}$ values should not be considered representative of the ULIRG population.

The estimated line ratios are, overall, in agreement with those reported by \cite{Papadopoulos+12a} for a sample of 70 local (U)LIRGs. Compared to other galaxy populations, such as high-$z$ SMGs \citep[see e.g.,][]{Bothwell+13,Spilker+14} and {\it BzK} galaxies \citep[][]{Daddi+15}, our ULIRGs show higher CO ratios. For a sample of main sequence galaxies at $z\sim1.25$, \cite{Valentino+20} measured\footnote{The luminosity ratios reported in their work are computed with respect to the CO(2--1) line luminosity, $r_{J2}\equiv L'_{{\rm CO}(J\rightarrow J-1)}/L'_{\rm CO(2-1)}$, as they are lacking CO(1--0) data.} $\langle r_{42} \rangle = 0.36$, $\langle r_{52} \rangle = 0.28$ and $\langle r_{72} \rangle = 0.09$, which are lower than the ones we measure for our sample of ULIRGs ($r_{42}= 0.50$, $r_{52}= 0.38$, and $r_{72}= 0.53$).
However, for galaxies above the MS (at $z\sim1.25$), classified as SBs and extreme SBs, they obtain higher values, closer to those measured in our work.
We warn the reader not to compare these ratios directly, as the work by \cite{Valentino+20} has low statistics for CO(4--3) detections (four measurements), and our sample has low statistics for CO(5--4) and CO(7--6), and so, the values may be subject to biases.
Even so, such comparison gives, at the very least, an idea of the extreme physical conditions of the molecular gas in local ULIRGs, with some presenting CO line ratios that resemble galaxies classified as SBs at high $z$.

Sources IRAS~10378+1109 and IRAS~16090-0139 present some of the highest excitation SLEDs. 
Indeed, these two sources (out of three in our sample) with CO(7--6) detections, show a remarkable high flux in this transition and are the ones responsible for driving up the average $r_{71}$ ratio value.
Both galaxies have interesting features in their observed spectra (Figures~\ref{fig:spec10378} and \ref{fig:spec16090}), for which the CO(7--6) emission seems to be broader than their corresponding lower-$J$ CO lines. In particular, the CO(7--6) emission from IRAS~10378+1109 appears to be red-shifted from the systemic velocity inferred from the low-$J$ CO lines by $\delta v\sim300$ ~\kms, whereas no other source in the sample exhibits a similar behavior\footnote{We checked the literature, and ALMA archival data for this galaxy, and it does not show evidence to be a particularly perturbed source.}.
While there are galaxies for which we observe line wings becoming more prominent with increasing $J_{\rm up}$, as can be seen, for instance, in IRAS~06035-7102 in Figure~\ref{fig:spec06035}, in those cases the appearance of more prominent wings does not significantly affect the centroid of the line. 
The apparent decrease of CO(7--6) flux in the systemic velocity component of IRAS~10378+1109 is indeed puzzling and deserves further inspection, ideally with spatially resolved data that allow for a kinematical analysis of the gas distribution in this galaxy. 
Interestingly, the same source also shows a remarkable trend of nearly thermalized CO emission up to $J_{\rm up}=7$, indicating there is a large fraction of gas in this galaxy under extreme physical conditions, and a likely continuous injection of energy into the medium in order to significantly populate the high-$J$ levels of the CO molecule.
In Section \ref{sec:link_exc_out} we discuss possible explanations for the observed high excitation in some of the ULIRGs in our sample.

\begin{figure}[tbp]
        \centering
        \includegraphics[width=.48\textwidth]{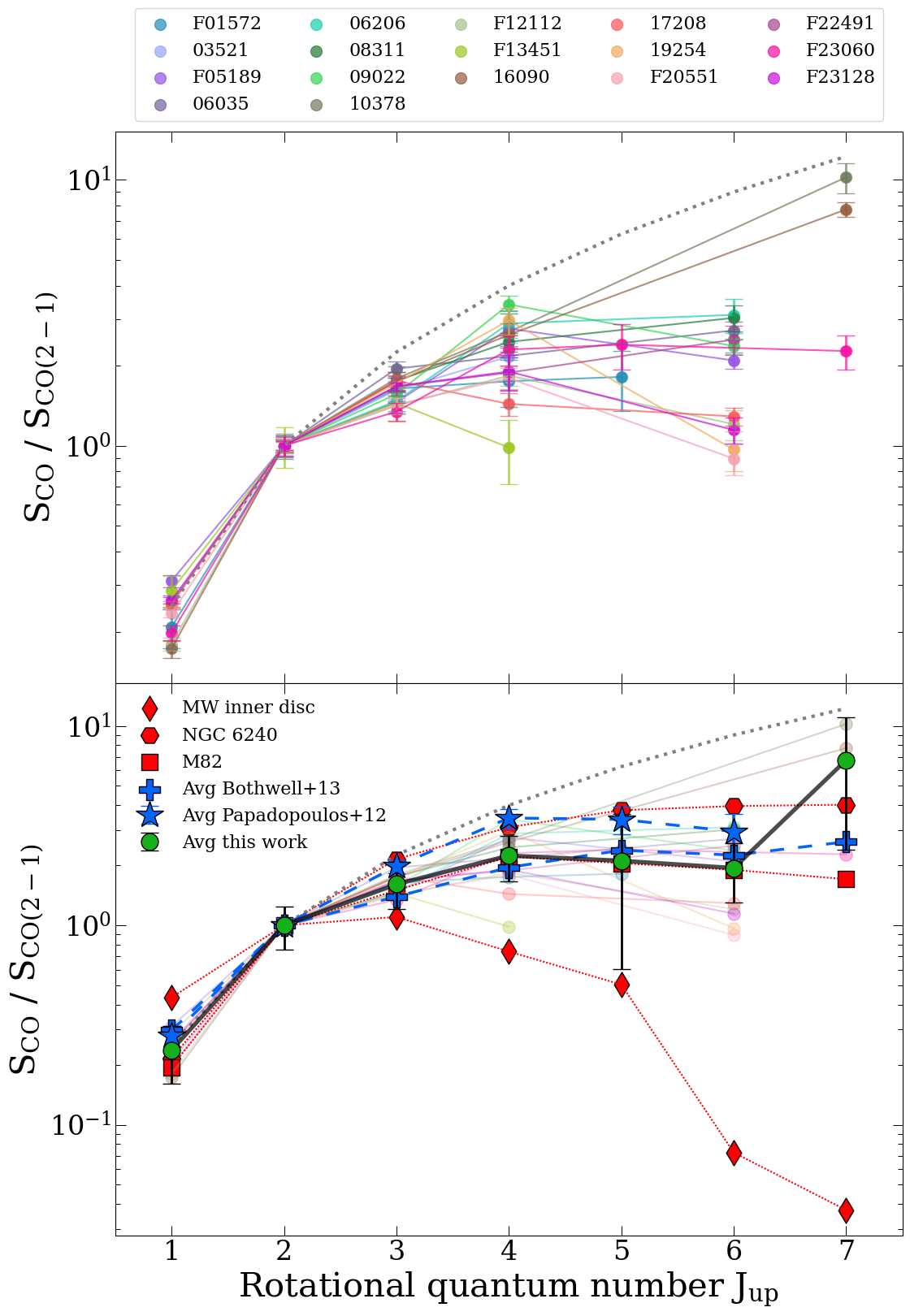}
        \caption{{\it Top panel:}  Observed CO SLEDs for the ULIRGs in our sample. The color coding is shown in the legend on top of the figure.
        {\it Bottom panel:} We compute the average CO SLED of our sample, and compare it to other well-studied galaxies, the starburst galaxy M82 \citep[][]{Mashian+15}, the nearby ULIRG NGC 6240 \citep[hosting a dual AGN and a SB;][]{Papadopoulos+14}, and the Milky Way inner disc \citep[][]{Fixsen+99}.
        We also show the average CO SLEDs for a sample of local (U)LIRGs studied by \cite{Papadopoulos+12a}, and a sample of high-$z$ SMGs \citep{Bothwell+13}.
        In both panels, the dotted gray line corresponds to a constant brightness temperature in the Rayleigh-Jeans regime (i.e., $S_{\nu}\sim\nu^2$), which is the trend expected for thermally excited CO emission{, in the optically thick limit for all lines}.
        All line fluxes are in units of Jy \kms, normalized by the {CO(2--1) flux measured by APEX}.}
        \label{fig:global_CO_SLEDs}
\end{figure}

The relative variations between the SLEDs of our ULIRGs sample are shown {in the upper panel of Figure~\ref{fig:global_CO_SLEDs}}, where we show all CO ladders in the same plot, normalized by the CO(2--1) velocity-integrated fluxes. 
By plotting all the sources together, the large variation in gas excitation within our sample, as traced by the CO($J>3$) lines, {becomes more evident}.
In the bottom panel of Figure~\ref{fig:global_CO_SLEDs} we show the average CO SLED of our sample (green circles), and compare it to other well-studied systems and galaxy populations. In particular, we show
the average CO SLEDs measured by \cite{Papadopoulos+12a} in local (U)LIRGs (blue stars in Figure~\ref{fig:global_CO_SLEDs}), and the one derived by \cite{Bothwell+13} for SMGs at $z=2.2$ (blue crosses in Figure~\ref{fig:global_CO_SLEDs}).
Our sample is about four times smaller than that studied by \cite{Papadopoulos+12a}, and hence our measured error bars are larger than those derived in their work. On the other hand, our data are more sensitive, can rely on broader bandwidths thanks to upgraded receivers, and are more uniform in terms of instrumental setups and analysis.
Both \cite{Bothwell+13} and our survey can rely on only a few measurements of the CO(7--6) transition, with our ratio being dominated by two high excitation sources, and as such, affected by a large difference and uncertainty in the estimated mean value.

While we are aware of the extreme nature of the ISM in ULIRGs, the difference between the CO SLEDs of our sample and the Milky Way inner disc \citep{Fixsen+99} is still remarkable, as shown by the bottom plot of Figure~\ref{fig:global_CO_SLEDs}.
In the Milky Way and in other disc galaxies, the H$_2$ gas excitation is dominated by much more quiescent gas, and so the CO SLED peaks at very low-$J$ CO rotational transitions.
The CO ladders of our sources are more in agreement with those measured in more extreme SBs, such as M~82 \citep[the archetypal SB galaxy, see][]{Mashian+15}, and NGC~6240 \citep[a known ULIRG to host both SB and AGN, see][]{Papadopoulos+14}.
Nevertheless, while these two galaxies may show similar gas excitation up to $J_{\rm up}=7$, they differ strongly for higher rotational lines, as it was evidenced in the study performed by \cite{Mashian+15}, analyzing {\it Herschel} CO data in the range $J_{\rm up} = 14-30$. 
These authors found that the CO ladders of M 82 and NGC 6240, albeit similar up to $J_{\rm up}\approx 8$, differ strongly at higher-$J$, with the former peaking around $J_{\rm up}\sim8$ and then quickly declining with increasing $J_{\rm up}$, as opposed to NGC~6240 that shows strong emission lines even up to $J_{\rm up}=28$.
This is a strong indicator of the need for CO($J_{\rm up}>8$) lines to properly constrain the physical properties of the gas in SBs and (U)LIRGs, which will be evidenced also by our analysis. 


\subsection{Lower and higher excitation sources}\label{sec:low_high_excitation_classif}

\begin{figure*}[tbp]
        \centering
        \includegraphics[width=.95\textwidth]{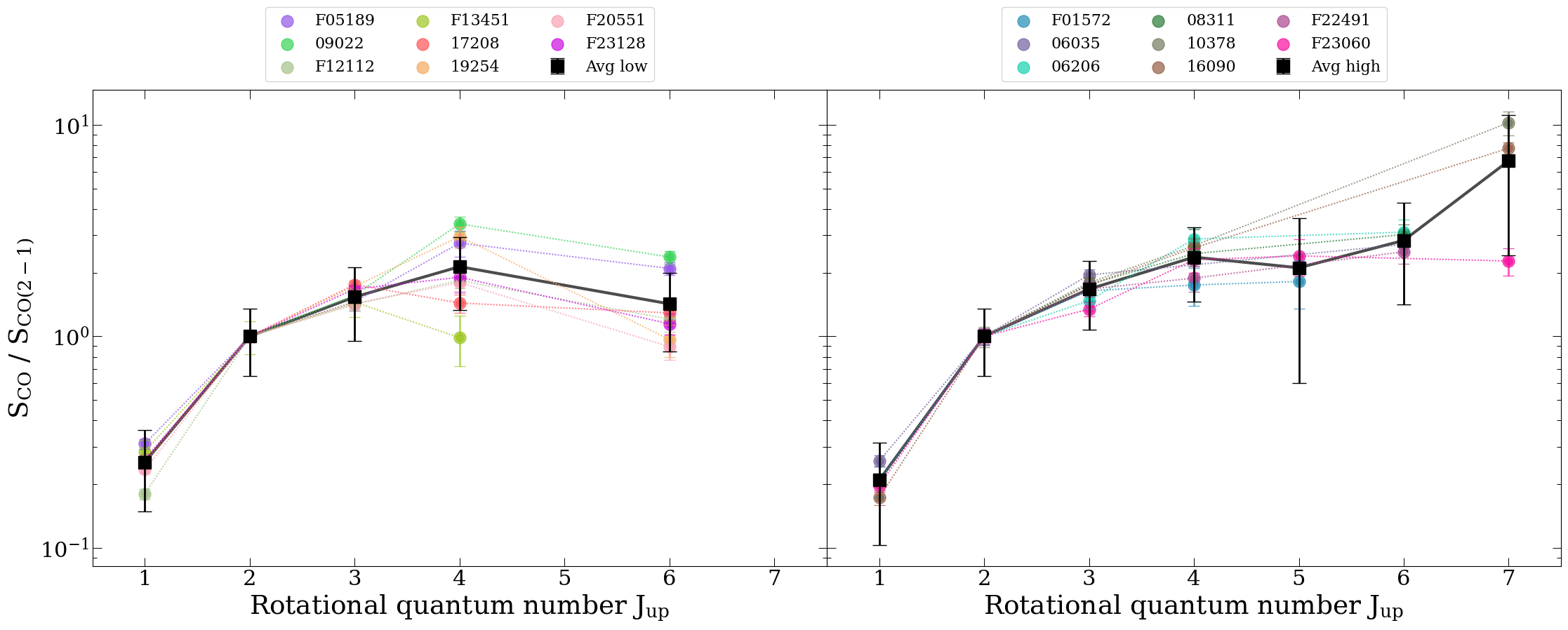}
        \caption{CO SLEDs of the ULIRGs classified as lower-excitation ({\it left panel}) and higher-excitation ({\it right panel}), following the classification criteria described in Section~\ref{sec:low_high_excitation_classif}. Each source is shown in a different color, according to the legend reported on the top of each panel.
        The average CO SLED of each group of galaxies is also computed and plotted using black square symbols connected by a black solid line}.
        \label{fig:low_high_CO_SLEDs}
\end{figure*}

To better visualize the different CO SLED shapes in our sample, we classify them into lower excitation and higher excitation\footnote{We choose the terms `lower-' and `higher-'excitation as opposed to `low-' and `high-'excitation, as ULIRGs still show significantly higher gas excitation than normal disc galaxies.}. 
We define lower-excitation sources those for which we observe a turn-over point in the CO SLED up to the observed rotational transition, and higher-excitation those for which we observe a plateau or increasing flux up to the observed CO transition.
There is only one source for which such classification is not possible, i.e., IRAS 03521+0028, for which we only have data up to CO(4--3) and no detection of a clear peak, leaving the shape of the CO ladder highly uncertain.
Figure~\ref{fig:low_high_CO_SLEDs} shows the CO SLEDs of lower-excitation (left panel) and higher-excitation (right panel) sources.
The CO(6--5) transition appears to be the key emission line to identify the different trends of CO ladders.
Indeed, the lower-excitation sources show an average CO SLED that peaks around $J_{\rm up}\sim 3,4$, with flux ratios decreasing starting from $J_{\rm up}= 6$.
We note that these sources, albeit classified as ``lower-excitation'' in this work, have higher excitation CO SLEDs than those measured in Milky Way-like galaxies and typical star forming discs \citep{Fixsen+99,Dannerbauer+09}. The average ratios of the ``lower-excitation'' ULIRGs are $\langle r_{31}\rangle_{\rm low~ex}=1.01$, $\langle r_{41}\rangle_{\rm low~ex}=0.45$ and $\langle r_{61}\rangle_{\rm low~ex}=0.15$.

Conversely, the higher-excitation ULIRGs show {peaks beyond the CO(4--3) transition}.
The average CO SLED derived for these sources shows a first plateau at $J_{\rm up}=4$, similar to the peak of the lower-excitation ones, and an unconstrained peak at higher $J_{\rm up}$, which suggests the need of at least two components to reproduce the observed SLEDs.
This apparent double-peaked CO ladder shape is driven, however, by the CO lines with lowest statistics in our sample, CO(5--4) and CO(7--6), for which we only have two and three measurements each, respectively, and so larger error bars; as such, the derived average CO SLED is subject to bias.
It is, however, constructive to discuss the CO ladder trends observed in these sources, as it is not the first time that they have been detected in the ULIRG population.
Indeed, it has been pointed out in previous studies that active SF galaxies likely host two (or even three) different molecular gas excitation components, one dominating the low-$J$ CO line emission, tracing the quiescent gas in a more diffuse and cold phase, and a second, more excited component that traces more extreme environments than are associated to AGNs (XDRs) and/or shocks \citep{Papadopoulos+14, Pereira-Santaella+14, Kirkpatrick+19}.
Such energetic processes are not uncommon in local ULIRGs, most of which are major galaxy mergers, where accelerated star formation and possibly feeding onto one or multiple SMBHs can be triggered by the gravitational interactions. In turn, the powerful radiative and mechanical feedback mechanisms resulting from the enhanced SMBH and SB activity can launch fast outflows embedding large amounts of molecular gas (up to $\sim10^{10}~M_{\odot}$ in the most extreme cases such as NGC~6240, see \citealt{Cicone+18}), hence possibly affecting the global excitation conditions of the molecular ISM, {which is the hypothesis that we aim to test in this work}. The connection between CO excitation and massive molecular outflows will be further explored and discussed in Section~\ref{sec:link_exc_out}.


\subsection{LVG modeling}\label{sec:lvg}

\begin{table}[tbp]
        \centering 
        \caption{Molecular gas properties derived from the single-component MCMC fit to the CO SLEDs of our sample of ULIRGs}.\label{table:single_MCMC}
        \resizebox{0.48\textwidth}{!}{%
        \begin{tabular}{lcccccc}
                \hline
                \hline
                Source      &   \multicolumn{2}{c}{$\log(n_{\rm H_{2}})$}   &    \multicolumn{2}{c}{$\log(T_{\rm kin})$}      &     \multicolumn{2}{c}{$\log(N_{\rm CO}/{\rm d}v)$}     \\
                            &   \multicolumn{2}{c}{$\log({\rm cm^{-3}})$}   &       \multicolumn{2}{c}{$\log({\rm K})$}       &  \multicolumn{2}{c}{$\log({\rm cm^{-2}\,km^{-1}\,s})$}  \\
                            & ${\rm med_{\pm1\sigma}}$ & ${\rm max_{post}}$ & ${\rm med_{\pm1\sigma}}$ & ${\rm max_{post}}$   &   ${\rm med_{\pm1\sigma}}$    &  ${\rm max_{post}}$     \\
                \hline
                F01572      &  $2.74^{+0.53}_{-0.43}$  &       $2.65$       &  $2.10^{+0.62}_{-0.56}$  &        $2.70$        &   $16.78^{+0.79}_{-0.45}$     &       $16.40$           \\ 
                03521       &  $2.58^{+0.79}_{-0.71}$  &       $2.34$       &  $1.88^{+0.68}_{-0.53}$  &        $2.43$        &   $17.14^{+0.86}_{-0.49}$     &       $17.13$           \\ 
                F05189      &  $2.05^{+0.12}_{-0.08}$  &       $2.01$       &  $2.91^{+0.07}_{-0.15}$  &        $2.97$        &   $17.20^{+0.07}_{-0.06}$     &       $17.17$           \\ 
                06035       &  $3.28^{+1.01}_{-0.85}$  &       $3.45$       &  $1.43^{+0.16}_{-0.12}$  &        $1.47$        &   $17.94^{+0.78}_{-0.68}$     &       $17.78$           \\ 
                06206       &  $5.85^{+1.13}_{-1.43}$  &       $6.50$       &  $1.17^{+0.08}_{-0.06}$  &        $1.13$        &   $18.71^{+0.56}_{-0.82}$     &       $19.12$           \\ 
                08311       &  $2.63^{+0.42}_{-0.27}$  &       $2.37$       &  $2.64^{+0.26}_{-0.51}$  &        $2.90$        &   $17.23^{+0.15}_{-0.19}$     &       $17.26$           \\ 
                09022       &  $3.02^{+0.95}_{-0.63}$  &       $2.63$       &  $1.69^{+0.68}_{-0.39}$  &        $2.36$        &   $17.41^{+0.98}_{-0.19}$     &       $17.39$           \\ 
                10378       &  $5.75^{+1.22}_{-1.51}$  &       $4.35$       &  $2.05^{+0.35}_{-0.27}$  &        $2.23$        &   $18.81^{+0.49}_{-0.71}$     &       $18.32$           \\ 
                F12112      &  $2.67^{+0.11}_{-0.07}$  &       $2.62$       &  $2.85^{+0.11}_{-0.23}$  &        $2.96$        &   $16.26^{+0.27}_{-0.27}$     &       $16.15$           \\ 
                F13451      &  $2.41^{+0.89}_{-0.67}$  &       $2.21$       &  $1.43^{+0.63}_{-0.36}$  &        $1.67$        &   $17.15^{+0.76}_{-0.67}$     &       $17.19$           \\ 
                16090       &  $2.73^{+2.17}_{-0.54}$  &       $4.67$       &  $2.08^{+0.21}_{-0.31}$  &        $1.81$        &   $18.80^{+0.38}_{-1.38}$     &       $17.43$           \\ 
                17208       &  $2.30^{+0.08}_{-0.08}$  &       $2.28$       &  $2.92^{+0.06}_{-0.14}$  &        $2.98$        &   $16.62^{+0.17}_{-0.17}$     &       $16.52$           \\ 
                19254       &  $4.40^{+0.90}_{-0.68}$  &       $4.59$       &  $1.45^{+0.28}_{-0.14}$  &        $1.46$        &   $16.15^{+0.68}_{-0.43}$     &       $16.14$           \\ 
                F20551      &  $2.47^{+0.12}_{-0.09}$  &       $2.40$       &  $2.78^{+0.15}_{-0.22}$  &        $2.92$        &   $16.37^{+0.20}_{-0.20}$     &       $16.27$           \\ 
                F22491      &  $2.54^{+0.39}_{-0.26}$  &       $2.30$       &  $2.70^{+0.22}_{-0.46}$  &        $2.92$        &   $17.16^{+0.15}_{-0.19}$     &       $17.17$           \\ 
                F23060      &  $4.96^{+1.58}_{-1.29}$  &       $4.07$       &  $1.26^{+0.07}_{-0.06}$  &        $1.26$        &   $18.43^{+0.78}_{-0.79}$     &       $18.88$           \\ 
                F23128      &  $2.34^{+0.17}_{-0.13}$  &       $2.25$       &  $2.62^{+0.25}_{-0.28}$  &        $2.85$        &   $16.89^{+0.20}_{-0.19}$     &       $16.73$           \\ 
                \hline
        \end{tabular}}

        \tablefoot{The values quoted for each parameter are the median with its corresponding $\pm1\sigma$ uncertainties of the marginal probability distribution functions (${\rm med_{\pm1\sigma}}$), and the maximum probability of the posterior distribution within the $\pm1\sigma$ range (${\rm max_{post}}$). All sources are modeled with a single excitation component {to avoid overfitting and minimise degeneracies, see main text for additional explanations for this choice.}}

\end{table}

%
%

{The focus of our work is on the relation between CO excitation and CO spectral broadening and asymmetries possibly tracing massive molecular outflows, which will be explored in Section~\ref{sec:link_exc_out}. However, in the following we report for completeness an analysis of the global CO ladders performed with }
\texttt{RADEX} \citep[see][]{vandertak+07}, a {non-LTE (local thermodynamic equilibrium) radiative transfer code, that uses an escape probability of $\beta=\left(1-e^{-\tau}\right)/\tau$} (where $\tau$ is the optical depth of the line), derived from an expanding sphere geometry, to compute the molecular line intensities.
{This geometry, similar to LVG codes, works} under the assumption that turbulent motions within star-forming clouds lead to large velocity gradients, and thus the photon escape probabilities are computed {accordingly} \citep{Sobolev+57,Ossenkopf+97}.
These are commonly used to model molecular line intensities, hence gas excitation {and SLEDs}, in galaxies as they can predict the shape of the rotational ladder by computing the collisional excitation of molecules (CO in our case) for a range of physical conditions.
In particular, we use the code developed by \cite{Yang+17}, \texttt{RADEX\_EMCEE}\footnote{https://github.com/yangcht/radex\_emcee}, which combines \texttt{pyradex}\footnote{Python wrapper of \texttt{RADEX} written by A. Ginsburg.} and \texttt{emcee}\footnote{Python implementation of the affine-invariant ensemble sampler for MCMC \citep{emcee}.} to model the CO fluxes, and find the best set of physical parameters that reproduce the SLEDs of our sample.
The CO collisional data are taken from the Leiden Atomic and Molecular Database (LAMDA) \citep{Shoier+05}. Following \cite{Yang+10}, we assume that H$_2$ is the only collisional partner of CO with an ortho-to-para ratio of 3.
The input parameters of the code are: the H$_2$ volume density ($n_{\rm H_2}$), the kinetic temperature of the molecular gas ($T_{\rm kin}$), the CO column density ($N_{\rm CO}$), and the size of the emitting region (or solid angle of the source).
{The results for the relative populations in the different rotational levels (i.e., the excitation) will depend on $N_{\rm CO}/{\rm d}v$, while line fluxes will scale as $\sim{\rm d}v$.}
The fluxes of {all CO lines} are equally scaled by the solid angle of the source, and as such the resulting shape of the CO SLED does not depend on this parameter and only acts as a normalization factor, hence only the first three parameters $\left(n_{\rm H_2},\, T_{\rm kin},\, N_{\rm CO}/{\rm d}v \right)$ are relevant. 

Instead of using \texttt{RADEX} to generate a grid of line fluxes for a range of input parameters to then fit the observed CO SLEDs, the code developed by \cite{Yang+17} adopts a Bayesian approach for fitting the observed CO fluxes. 
A Markov chain Monte Carlo (MCMC) calculation is performed to explore the continuous parameter space and it calls \texttt{pyradex} in each iteration for computing the CO fluxes for a given combination of $\left\{ n_{\rm H_2},\, T_{\rm kin},\, N_{\rm CO}/{\rm d}v \right\}$, it then samples the posterior probability distribution function to obtain the next most probable set of parameters to use as inputs.
This approach allows for a better sampling in the parameter space while avoiding unnecessary calculations, thus resulting in a faster convergence.
We used a flat log-prior for all the parameters to set boundaries for the parameter space. 
The ranges are as follows: volume density $n_{\rm H_2}$ from $10^{1.5}$ to $10^{7.5}$ cm$^{-3}$, kinetic temperature $T_{\rm kin}$ from the CMB temperature at the redshift of the source ($T_{\rm CMB}=2.7315(1+z)$) to $10^3$ K, and CO column density per unit velocity gradient $N_{\rm CO}/{\rm d}v$ from $10^{14.5}$ to $10^{19.5}$ cm$^{-2}$km$^{-1}$s.
Additionally, we also use a range for the velocity gradient ${\rm d}v/{\rm d}r=0.1-1000$ \kms~{pc$^{-1}$} \citep[see][]{Tunnard+16}, which sets a limit for the ratio between $N_{\rm CO}/{\rm d}v$ and $n_{\rm H_2}$\footnote{A fixed CO abundance per velocity gradient $X_{\rm CO}/({\rm d}v/{\rm d}r)=10^{-5}${~km$^{-1}$ s pc} is adopted \citep{Weiss+07,Liu+21}, and thus a range in velocity gradients, constraints the ratio between $N_{\rm CO}/{\rm d}v$ and $n_{\rm H_2}$, via $N_{\rm CO} = X_{\rm CO}\,n_{\rm H_2}\,{\rm d}v/({\rm d}v/{\rm d}r)$.}.
The prior probabilities that fall outside of the boundaries are set to 0. 

As discussed in Section \ref{sec:sleds}, there is a wide range of CO excitation conditions within the ULIRG population, and while previous studies show that the CO emission is likely dominated by two excitation components, the lack of CO($J_{\rm up}>7$) lines in our study prevents us from fitting two different components without being subject to significant degeneracy between the derived gas properties.
{For these reasons, in this work}, we are fitting the observed CO SLEDs with a single-component LVG model, approximating all CO emission to be characterized by a single kinetic temperature, H$_2$ volume density, and column density per velocity gradient.
The modeled CO SLEDs from the best-fitting estimates of $\left(n_{\rm H_2},\, T_{\rm kin},\, N_{\rm CO}/{\rm d}v \right)$ and their $\pm1\sigma$ uncertainties, are shown in Figures \ref{fig:specF01572} through \ref{fig:specF23128}, together with our CO flux measurements. 

The LVG modeling results are summarized in Table~\ref{table:single_MCMC}.
We obtain best-fit values in the following ranges: $n_{\rm H_2} \approx 10^{2.0}-10^{5.8}$ cm$^{-3}$, $T_{\rm kin}\approx10^{1.2}-10^{2.9}$ K, and $N_{\rm CO}/{\rm d}v\approx10^{16.2}-10^{18.8}$ cm$^{-2}$ (km~s$^{-1}$)$^{-1}$.
These are overall consistent with the results of single-component LVG modeling of other samples of local ULIRGs \citep[see e.g.,][]{Ao+08, Weiss+05,Rosenberg+15} and high-$z$ SMGs \citep[e.g.,][]{Canameras+18,Yang+17,Spilker+14}.
However, for the higher-excitation sources, driven by the high mid-$J$ CO line fluxes, we derive extreme conditions that are physically improbable to be representative of the average state of the ISM in these galaxies.
For instance, the high fluxes measured in the CO($J_{\rm up}=4,6$) lines for IRAS~06206-6315 boost the average H$_2$ volume density of the galaxy to $\log{\left(n_{\rm H_2}\,{\rm [cm^{-3}]}\right)}\approx5.6^{+1.1}_{-1.4}$, and a similar effect is observed in IRAS F20551-4250, though in this latter case, the parameter boosted to higher value is the gas kinetic temperature to $\log{\left(T_{\rm kin}\,{\rm[K]}\right)}\approx2.8^{+0.2}_{-0.2}$.
These values are indeed highly uncertain, as can be evidenced by the $\pm1\sigma$ uncertainties, or by the posterior probability distributions showed in Appendix \ref{app:spec_sleds}, where it becomes clear that the parameter space is highly biased to higher values, and so, overall poorly sampled. This effect is seen in IRAS F05189-2524, IRAS F12112+0305, IRAS 17208-0014, IRAS F20551-4250, and IRAS F23128-5919.
The reason for this roots from the Ansatz of using a single LVG model.
For instance, the CO ladder of IRAS 17208-0014 does not fall down steeply after its peak, but remains significantly excited.
If we were to overplot a single LVG model with reasonable temperature (e.g., $50$~K) that reproduces the data up to the peak at $J_{\rm up}=4$, it would drop much faster for transitions above peak compared to the measurement.
This can easily be captured by adding a second component with a higher density that broadens the CO SLED. However, for our single component approach, the temperature has to {reach} very high values in order to fit the $J_{\rm up}=6$ line.
A similar effect will be at play for those solutions with very high density (e.g., IRAS 06206-6315).
In such cases, the CO SLED is not well sampled beyond the peak, thus the shape of the drop-off at higher-$J$ CO lines remains unconstrained.
The CO fluxes can then be reproduced with a low-temperature model, however, a correspondingly high density is needed to move the peak of the CO SLED to the highest $J_{\rm up}$ measurement. 
As a result, changing the flat log-priors in the MCMC to cover a wider range in the sampling of the parameter space, does not lead to better results, as the outcome still leads to unphysical high values of volume density and/or kinetic temperature in order to reproduce the high fluxes measured for higher-$J$ CO lines.
These results exemplify the degeneracy between the $n_{\rm H_2}$ and $T_{\rm kin}$ on the shape of the CO SLED, as an increase in either will lead to more excited gas. 
In order to break such degeneracy, we need complementary {high resolution} CO($J_{\rm up}>7$) line data.

A single-component LVG fit 
is certainly an over simplification, and provides only a crude approximation to the physical properties that characterize the entirety of the ISM within these sources. 
The CO SLED plots reported in Appendix~\ref{app:spec_sleds} confirm that the LVG single component modeling is not adequate for some of our sources.
In particular, the sources mentioned above with poor sampling of the posterior probability distribution, as well as IRAS 06206-6315, IRAS 09022-3615, IRAS F22491-1808 and IRAS F23060+0505, are some of the galaxies that likely require a second excitation component.
Many of these sources have already been studied with {\it Herschel} across the much broader range of CO transitions.
\cite{Pearson+16} grouped the SLEDs of their ULIRG sample into three categories: flat, increasing fluxes (low- to high-$J$), and decreasing fluxes. These authors found that the SLED's slopes correlate to the FIR colors as traced by the $60\mu$m/$100\mu$m color index (lower values corresponding to colder sources), with warmer ULIRGs showing higher excitation CO SLEDs \citep[a trend already seen in studies by][]{Rosenberg+15,Lu+14}.
We also explored a possible correlation between the lower- and higher-excitation ULIRGs in our sample and their FIR color (using IRAS measured $60\mu$m and $100\mu${m} fluxes), and we did not find a clear correlation of higher excitation for warmer ULIRGs.
However, this may be due to the degeneracy caused from classifying sources into lower- and higher-excitation based on CO SLEDs that are sampled only up to CO(7--6) in our case. And indeed, some of the sources belonging to the `lower-excitation' group under our criteria, have been observed to have high-excitation CO SLEDs when probing CO($J>7$) lines {(e.g., IRAS 09022-3615, IRAS F05189-2524, IRAS 17208-0014)}. 

\cite{Pearson+16} showed that, while there are sources for which a single-component CO SLED fit is sufficient, the CO ladder of IRAS~09022-3615 (classified as a warm ULIRG in their study, and included also in our work) requires three excitation components in order to reproduce the observed CO fluxes up to $J_{\rm up}=13$: (i) a cold (47~K) component peaking at $J_{\rm up}\sim3$, (ii) a warm (205 K) component peaking at $J_{\rm up}\sim6$, and (iii) a third hot (415 K) component that dominates at high-$J$ transitions, peaking around $J_{\rm up}\sim11$.
They propose that the hottest component traces the inner PDRs, {shocks and/or AGN-heated gas (XDRs)}.
In our analysis, the best-fit parameters derived for IRAS 09022-3615 from the single LVG component model are $n_{\rm H_2} \approx 10^{2.6}$ cm$^{-3}$ and $T_{\rm kin}\approx250$ K. This result is in agreement with the warm component (ii) of \cite{Pearson+16}, which dominates the mid-$J$ CO emission.
Nevertheless, this source is classified as a lower-excitation ULIRG under the criteria used in our work.
This is due to the lack of CO($J\geq7$) lines that would evidence the high excitation nature of this source, demonstrating once more the need to observe higher-$J$ CO lines to properly determine the physical conditions of galaxies.

Our sample overlaps also by two sources with that studied by \cite{Rosenberg+15}. These authors classify the ULIRGs into three groups (similar to \cite{Pearson+16}) depending on the drop-off slope of the CO ladder from the $J = 5-4$ transition and on-wards. Their ``Class I'' objects have the steepest drop-offs, the ``Class II'' sources peak at around $J=6-5$, and ``Class III'' objects are characterized by flat CO SLEDs.
For Class II and III (the highly excited classes) sources, heating mechanisms besides UV heating are required to explain the high-$J$ CO emission.
Sources IRAS~17208-0014 and IRAS~F05189-2524 are classified as Class III objects in their study, while in this work they both lie in our lower-excitation group.
According to \cite{Pereira-Santaella+14}, shocks play an important role in setting the higher-$J$ CO excitation conditions of IRAS~F05189-2524.
Interestingly, in our analysis, the single-component LVG model does not provide a good fit to either of these two sources (see Figures \ref{fig:specF05189} and \ref{fig:spec17208}), with poorly sampled posterior probability distributions due to the bias caused by the high CO($J_{\rm up}=4,6$) fluxes, hence resulting in a derived kinetic temperature that is close to the upper boundary with misleading uncertainties.

Despite many of our results and several literature studies suggesting the need of at least two excitation components to properly reproduce the observed CO SLEDs of ULIRGs, and so derive accurate physical conditions of their ISM, we restrain ourselves from adding another component to our fitting procedure, as our ground-based CO data, lacking high-$J$ CO transitions, do not allow us to resolve the ambiguity between the kinetic temperature and the H$_2$ volume density in the LVG models.
Therefore, we caution against an over-interpretation of the specific best-fit parameters derived for individual sources reported in Table \ref{table:single_MCMC}, specially those classified as higher-excitation sources, for which the single-component model proves to be inadequate. On the other hand, thanks to the high S/N and spectral resolution, {our data are most adequate to perform} an analysis of the links between spectral line profiles and global CO excitation conditions, {which is not possible from the low spectral resolution {\it Herschel} spectra}. Section~\ref{sec:link_exc_out} will focus on exploring such link.


\section{A link between CO excitation and molecular outflows}\label{sec:link_exc_out}

\subsection{Analysis of line widths as a function of $J_{\rm up}$ of the CO transition}\label{sec:linewidths}

\begin{figure*}[tbp]
        \centering
        \includegraphics[width=.975\textwidth]{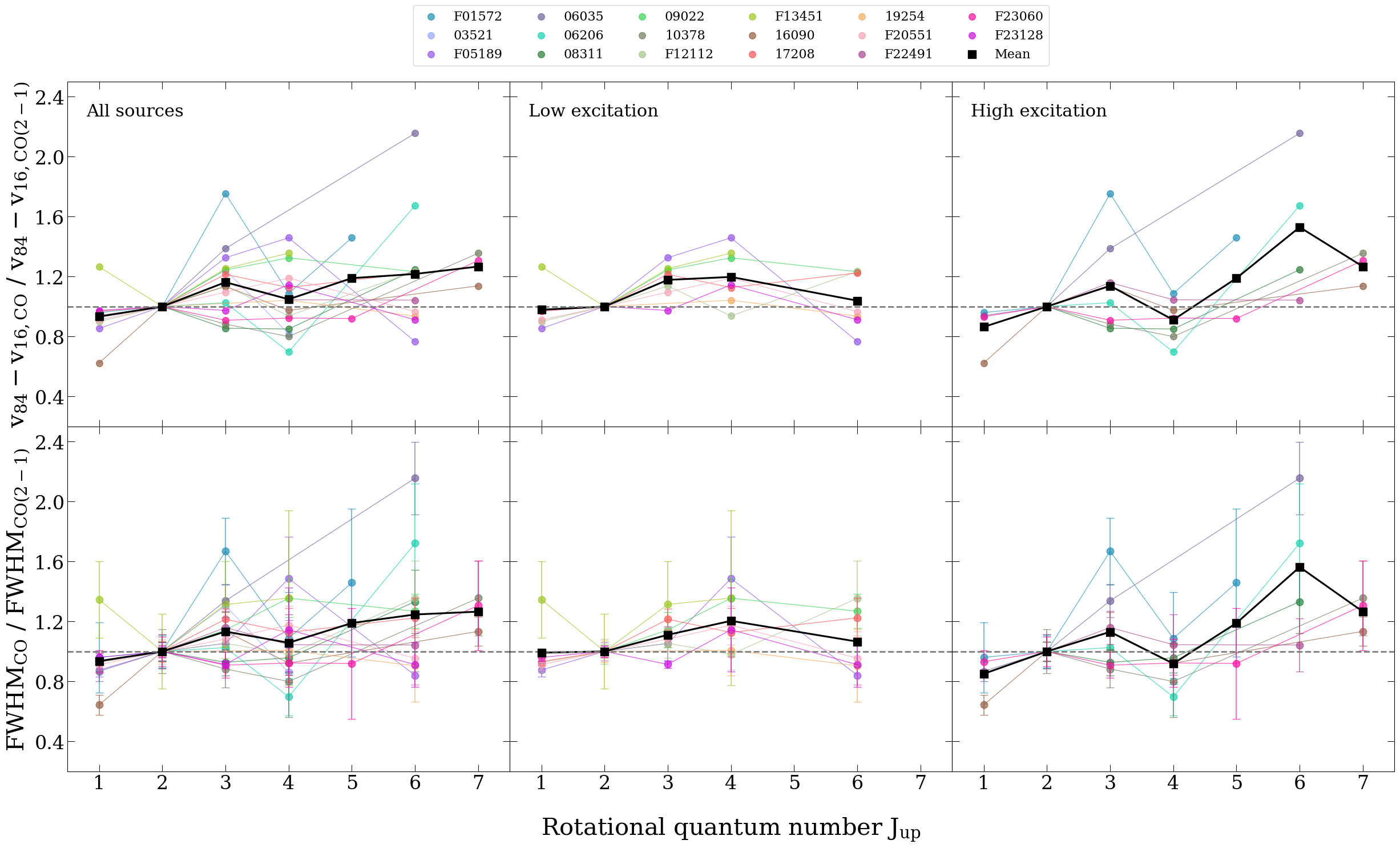}
        \caption{Observed CO line widths (normalized by the CO(2--1) line width) as a function of upper rotational quantum number $J_{\rm up}$. {\it Top row:} CO line widths computed using the velocity percentile ${\rm v_{84}}-{\rm v_{16}}$ interval, derived from the analytical expression of the best-fit line profile obtained with the multi-Gaussian fit. {\it Bottom row:} line widths obtained via a single Gaussian spectral fit to each CO emission line. {\it Left column}: all the sources in our sample; {\it middle} and {\it right columns} show the ULIRGs with low and high excitation CO SLEDs, respectively (see explanation in Section \ref{sec:sleds}). The sources are color coded as presented in the upper legend, the mean trend is computed for each plot and shown with black square markers.}
        \label{fig:all_linewidths_vs_j}
\end{figure*}

In the following, we will test the hypothesis of a relation between CO excitation and {the appearance of broad and/or high-velocity CO spectral components}, which may indicate that massive molecular outflows are responsible for setting the {\it global} CO excitation conditions of ULIRGs, an hypothesis already advanced in previous works but never confirmed or discarded \citep{Cicone+18, MontoyaArroyave+23}. 
We will start by exploring the presence of any trends between the observed CO line widths and the upper rotational quantum number $J_{\rm up}$.



Before showing the results, we discuss briefly what we expect from such analysis from a classical PDR perspective. Carbon monoxide is unique in being largely abundant and having low-$J$ transitions that are easily excited under mild conditions (i.e., low-density and low-temperature environments), causing their emission to be commonly optically thick \citep{Narayanan+14}. 
Conversely, mid- to high-$J$ CO lines require the gas to be in more extreme conditions (either denser or warmer environments) to be excited.
In the standard PDR scenario, such environments are typical of molecular cloud cores (i.e., high density gas) and/or active star-forming regions (high temperature and density) \citep{Hollenbach+Tielens+97}.
It follows that mid- and high-$J$ CO lines may arise from regions with different kinematics: (i) dense cores of GMCs, where the gas is more shielded from processes that introduce turbulence to (or even disrupt) the gas cloud; 
and (ii) regions with active star formation, injecting both kinetic and thermal energy into the surrounding medium, leading to a layer of gas with turbulent motions emitting higher-$J$ CO lines.
Indeed, emission coming from the latter may broaden the spectral line profile due to turbulence; however, the amount of gas found in such regions is overall a small fraction of the total molecular gas budget in galaxies, and as such we would expect the contribution to the total integrated line profile to be minimal.
In fact, in the classical PDR interpretation, high-$J$ CO lines are expected to depart from thermalisation \citep{Narayanan+14}, and are more likely to be optically thin, with line widths that are either narrower than the low-$J$ CO lines, or at most consistent if we assume that all the gas shares the same galaxy-wide kinematics.

Since our APEX single-dish spectra enclose the majority of the ISM of these sources (beam values ranging from $8-27$ arcsec, corresponding to $\sim 14-48$ kpc), any observed CO spectral line broadening will be dominated by macroscopic processes. {We rule out opacity \citep[see e.g.,][]{Hacar+16} to produce any measurable line broadening effects on such scales, as demonstrated by the near 1:1 correspondence observed between the line widths of low-$J$ CO lines and the optically thin atomic carbon lines measured in our sample, presented and discussed in \cite{MontoyaArroyave+23}}\footnote{{Deviations between CO and [CI] line widths are marginal but interesting on their own, as discussed in that paper, but the fact that they are only at the $\sim10$\% level despite the widely different expected optical depths of [CI] and low-$J$ CO lines, the fact that they affect the sources with the broadest spectra ($\sigma_{\rm CO}\gtrsim150$~\kms) and they are clearly evident only in the line wings of the spectra rather than in the core, strongly suggest a kinematic origin, rather than an opacity origin.}} Doppler broadening due to kpc-scale rotating molecular gas disks can {produce line wings up to v$\sim200-250$~\kms on each side of the line in galaxy-integrated spectra of ULIRGs}. In the presence of massive molecular outflows, the {total} observed CO line spectra can extend up to absolute line-of-sight velocity values of $\sim1000-1500$~\kms in {the most extreme cases} (see, e.g. \citealt{Cicone+14}). 
In addition, inflows, tidal tails, and merging companions included in the beam can also contribute to the observed CO line broadening. However, due to the extremely high incidence of massive molecular outflows observed in ULIRGs \citep[$>70\%$, see also OH outflow properties reported in Table 1 in][]{MontoyaArroyave+23},
it is reasonable to assume that high-velocity and broad CO line components in our sample {contain also outflowing material}.

The most direct measurement of CO line widths can be obtained through a single Gaussian fitting, by using the full width at half-maximum (FWHM). However, the total CO line profiles of local ULIRGs often
present broad wings, asymmetries and/or double-peaked emission. In these cases, a single Gaussian component may be a poor fit.
To account for this, we also estimate the 16th-84th percentile velocity interval (${\rm v_{84}}-{\rm v_{16}}$), derived from the analytical expression of the best-fit line profile obtained with the multi-Gaussian fit.
In Figure~\ref{fig:all_linewidths_vs_j} we show the CO line widths, estimated with both methods (top and bottom panels), plotted as a function of $J_{\rm up}$ of the transition. The left panels show the whole sample, while the middle and right panels display separately the lower- and higher-excitation sources, respectively, as classified in Section~\ref{sec:sleds}.
{The sample shows a high variance, however,} interestingly, when averaging the data (black solid line in the left panel of Fig.~\ref{fig:all_linewidths_vs_j}), we retrieve a {mean} increasing trend of CO line widths as a function of $J_{\rm up}$, with a Pearson correlation coefficient of 0.91 and a $p$-value of $4.3\times10^{-3}$ for the ${\rm v_{84}}-{\rm v_{16}}$ line widths measurements (for the FWHM approach we retrieve 0.95 and $1.2\times10^{-3}$, respectively), hence it is statistically significant at the $4\sigma$ level.
The relations obtained with the two line width estimate methods show remarkably similar results.
The middle panels of Figure~\ref{fig:all_linewidths_vs_j} show that for lower-excitation sources the line widths increase up to CO($J_{\rm up}\sim3,4$), similar to the shape of the average CO SLED retrieved for these sources.
For higher-excitation sources, the average trend shows increasing line widths up to the CO(6--5) transition, and a decrease for the CO(7--6) line (see e.g., Figures \ref{fig:spec06206} and \ref{fig:spec16090} for sources with broader high-$J$ CO lines).
The behaviors for different sources indicate that the trend of broader line profiles for higher-$J$ CO lines is consistent with being driven by the higher-excitation ULIRGs in our sample.
To better visualize the broadening observed from low- to high-$J$ CO transitions in the higher excitation ULIRGs, in Figures \ref{fig:specF01572} through \ref{fig:specF23128} we also show a plot overlaying the best-fit line profiles (normalized to their maximum value) obtained for all the available CO transitions for each source. {The sources showing the most remarkable line broadening in higher-$J$ CO transitions are: IRAS~06035-7102 (Figure~\ref{fig:spec06035}), IRAS~06206-6315 (Figure~\ref{fig:spec06206}), IRAS~08311-2459 (Figure~\ref{fig:spec08311}) and IRAS~16090-0139 (Figure~\ref{fig:spec16090}). }

{This result is remarkable considering that it is obtained from an analysis of single dish beam-integrated spectra, probing regions as large as $\sim10-15$~arcsec in the mid-$J$ CO transitions, corresponding to tens of kpc.}
As such, the processes at the origin of {the observed trends} must be predominant on galactic scales and affect a significant portion of the molecular ISM in these sources, {otherwise their signature would be highly diluted in the large beam}. 
{This result challenges classical nuclear ($r<1$~kpc) PDR or XDR scenarios for the origin of the high CO excitation in these ULIRGs, as already}, emerged from the results of the CO SLED analysis presented in Section~\ref{sec:lvg}.
 
Most previous galaxy-wide CO SLEDs surveys have been carried out with {\it Herschel} spectroscopic data \citep[see][]{Kamenetzky+14,Pearson+16,Liu+15,Greve+14,Lu+14,Pereira-Santaella+14,Mashian+15}, 
and, while such studies provide a global view of the extreme physical conditions of the molecular gas in (U)LIRGs, they lack the sensitivity and spectral resolution needed to perform a {spectral profile} analysis.
Consequently, only few literature studies have evaluated possible variations of the line widths of different CO rotational transitions. 
Most previous works found similar line widths throughout low- to high-$J$ CO lines, and usually opted for deriving an average line width for all CO transitions for individual sources \citep[e.g.][]{Yang+17,Greve+05,Weiss+07}.
In the study of 40 SMGs at redshift $z\sim 1.2-4.1$ performed by \cite{Bothwell+13}, they found no significant differences between the line widths of low-$J$ (i.e., $J = 2-1$) and higher-$J$ (i.e., $J_{\rm up} \geq 3$) lines, with the mean values of the distributions agreeing to within their $1\sigma$ errors.
From the FWHMs derived for each of their SMGs, \cite{Bothwell+13} found that ``hotter dust'' ULIRGs are kinematically similar to SMGs as traced by their line widths, with ULIRGs having on average narrower line profiles than SMGs. 
\cite{Boogaard+20} reached a similar conclusion based on a study of spectrally resolved CO($J_{\rm up}\leq8$) emission lines in a sample of SF galaxies at $0.5<z<3.6$.
Their measurements showed consistent widths for different CO rotational transitions.
They argued that the gas distribution does not differ strongly in different transitions, leading to almost constant line widths at different $J$. The galaxies in their sample are, however, less extreme than our ULIRGs, with lower infrared luminosities and SFR surface densities than the SMGs and SF galaxies studied in e.g. \cite{Bothwell+13,Spilker+14,Daddi+15}.

An interesting take on the diagnostic power of CO line widths is that of \cite{Rosenberg+15}, who argued that the width of an emission line could indicate more mechanical energy available in the host galaxy, and, as such, more heating of the molecular gas. 
In their study, these authors combined ground-based low-$J$ ($J\leq3$) CO observations\footnote{The ground-based data are collected with different single-dish telescopes and interferometers.}, with {\it Herschel} CO($4\geq J\geq13$) data for 29 (U)LIRGs.
They measured the FWHMs of the spectrally resolved CO(1--0) lines, and found that the lower-excitation sources display narrower CO(1--0) lines, while the higher-excitation galaxies span a wide range of CO(1--0) line widths.
They propose that, by combining the information from the
measured CO(1--0) line widths with the AGN luminosity, it is possible to shed light onto the origin of the high CO excitation. In particular, for the higher-excitation sources that cannot be explained only with UV heating from massive stars, these authors propose that broad lines may be indicative of mechanical heating, especially when the AGN contribution is low and so heating from X-rays and Cosmic Rays may not be important. 

According to \cite{Bournaud+15}, the high CO gas excitation measured in starburst mergers can be driven by strong turbulent compression motions associated with tidal interactions, which lead to high-density environments. This suggests that non-ordered motions in galaxy mergers cannot be neglected when studying molecular lines.
However, the simulation used by \cite{Bournaud+15} does not include any prescription for AGN feedback, which is important in ULIRGs, and, more importantly, it does not account for cold molecular gas embedded in galactic outflows. The latter is now a well-established result, thanks to the significant observational efforts made in the past 10 years \citep[see e.g.,][]{Veilleux+20}, and it is even more relevant in ULIRGs where massive molecular outflows are almost ubiquitous and can be as massive as to embed a significant portion of their molecular ISM \citep{Cicone+18}.

\subsection{A possible relation between CO excitation and molecular outflows}\label{sec:mol_outflows}

In this section we explore further the marginal trend of increasing line width in higher-$J$ transitions presented in Section~\ref{sec:linewidths}, with the aim of testing whether such relation can be linked more closely to massive molecular outflows. Since the positive relation observed in Figure~\ref{fig:all_linewidths_vs_j} is driven by the higher-excitation ULIRGs, here we focus only on these objects (8 out of the total sample of 17 ULIRGs).

Due to the turbulent nature of the ISM and the merging state of ULIRGs, it is not straightforward to link the broad/narrow components of {galaxy-integrated} CO line spectra to outflowing/non-outflowing gas, in absence of additional spatial information. Nevertheless, most of these ULIRGs are known to host powerful molecular outflows, which have been unambiguously detected through blue-shifted OH absorption lines {\citep{Spoon+13,Veilleux+13,Sturm+11}}.
Therefore, as already discussed in \cite{MontoyaArroyave+23}, where we performed a similar analysis limited to the three lowest-$J$ CO transitions, we can assume that, in first approximation, the broad and/or high-$v_{\rm cen}$  components of the CO spectra are most likely dominated by molecular gas embedded in outflows, while the low-$v_{\rm cen}$ and low-$\sigma_v$ components are most likely dominated by the gas in rotating disks and in other regions of the `non-outflowing' ISM. 
Following this line of reasoning, we performed a multi-Gaussian component simultaneous fit to all CO transitions available for each source, by tying $v_{\rm cen}$ and $\sigma_v$ of each Gaussian to be equal in all CO transitions, allowing only their amplitudes to vary freely in the fit \citep[for details see][]{MontoyaArroyave+23}. We allowed the fit to use up to three Gaussian functions for each line, in order to capture any possible asymmetries/wings in the observed spectra. Such simultaneous multi-Gaussian spectral fits, performed only for the higher-excitation sources, are displayed in Appendix~\ref{app:spec_sleds}.

\begin{figure}[tbp]
        \centering
        \includegraphics[width=.48\textwidth]{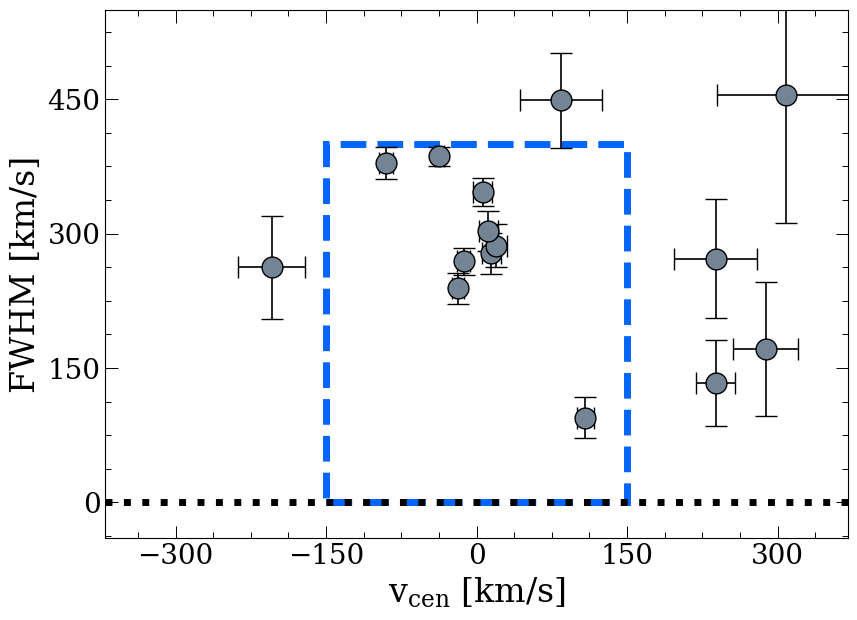}
        \caption{FWHM as a function of the central velocity of all Gaussian components employed in the simultaneous CO spectral fits performed on the higher-excitation sources. The blue dashed rectangle encloses the region of the parameter space that we ascribe to the quiescent gas. The components falling outside this regions are classified as being dominated by outflowing gas. }
        \label{fig:fwhm_vcen}
\end{figure}

The next step is to classify the different Gaussian components. To do so, we 
plot them in the FWHM-$v_{\rm cen}$ parameters space, shown in Figure~\ref{fig:fwhm_vcen}, including all eight higher-excitation sources.
The plot shows a clustering of spectral components with high S/N (small errorbars) at low-$v_{\rm cen}$ and low FWHM values, which are consistent with probing the dynamically quiescent (i.e. non-outflowing) gas. We then draw a region in the FWHM-$v_{\rm cen}$ space enclosed within $-150 < v_{\rm cen}\;{\rm [km\;s^{-1}]}<150$ and ${\rm FWHM}<400$ \kms (rectangle in Figure~\ref{fig:fwhm_vcen}), which delimits these components classified as quiescent. We classify the components lying outside this area as components containing outflows.
We note that a similar approach was used by \cite{Cicone+18} to identify the outflow-dominated molecular line emission from NGC~6240. {An important take-away point from \cite{Cicone+18} is that the broad wings are highly attenuated in the galaxy-integrated CO spectrum of NGC~6240, where the components at velocities above 500~\kms~show fluxes that are $<5$\% of the peak of the line, despite this source hosting one of the most massive and extended molecular outflows known. Indeed, the spatially resolved analysis of the CO data of NGC~6240, reported in the same paper, demonstrates that (i) along the east-west direction of expansion of the outflow, several kpc away from the known nuclear molecular disc(s) in NGC~6240 \citep{Medling+19}, the broad CO components can reach velocities beyond 1000~\kms, and at the same time still show a significant flux at low-to-zero line-of-sight velocity, and that (ii) as much as 60$\pm$20~\% of the total H$_2$ reservoir in the central $\rm 6~kpc\times3$~kpc region of NGC~6240 is embedded in the outflow. This means that the spectral signature of even the most extreme molecular outflows can be highly diluted in galaxy-integrated spectra of ULIRGs, where the detection of even faint (a few \% of the peak) high-$v$/broad components can indicate a potentially extreme outflow event. Such aperture-dilution effects are one of the reasons why the CO-integrated galaxy spectra (almost) never show velocities as high as the maximal velocities seen in blue-shifted OH absorption lines detected against the FIR continuum, the other reason being that the OH lines generally probe the innermost parts of these outflows, close to the launching base \citep[$<200$~pc for the high-velocity components of OH, see e.g.,][]{Gonzalez-Alfonso+17}.}
{Indeed, when exploring the relation between the OH outflow velocities \citep[reported in][]{Spoon+13,Veilleux+13} and gas excitation for our sample, we find no evidence of higher excitation for ULIRGs with measured larger OH outflow velocities. In fact, some sources classified as higher-excitation in our work, are only detected in absorption in OH, and thus lack outflow velocity measurements. This has been already discussed in \citet{MontoyaArroyave+23} for our parent sample of (U)LIRGs, where we argue that OH velocities are not representative of the strength of molecular outflows, as they depend on other factors (orientation of the outflow with respect to the Line of Sight, the presence and strength of the FIR continuum).}


\begin{figure}[tbp]
        \centering
        \includegraphics[width=.48\textwidth]{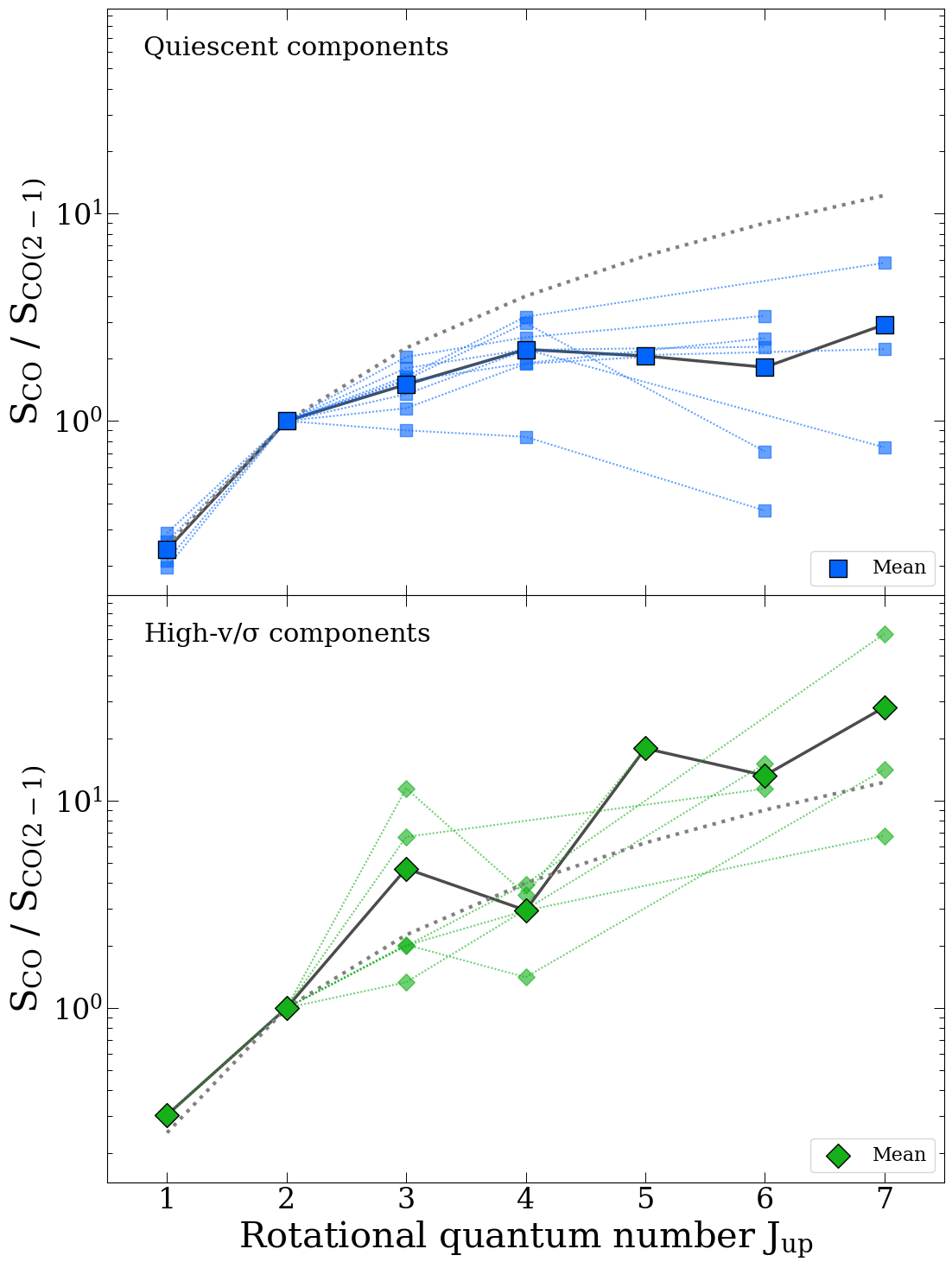}
        \caption{\textit{Top panel:} CO SLEDs for the spectral components that lie within the blue rectangle in Figure \ref{fig:individual_comp_SLEDs} probing dynamically quiescent gas.
        \textit{Bottom panel:} CO SLEDs for the spectral components that lie outside the blue rectangle in Figure \ref{fig:individual_comp_SLEDs}, likely containing molecular outflows in these sources.
        Only `higher-excitation' ULIRGs are included in this analysis (see Section~\ref{sec:low_high_excitation_classif}). Quiescent components show significantly less excited CO SLEDs compared to the high-$v$/high-$\sigma$ spectral components.
        {The dotted gray lines show the thermalized optically thick limit (in all lines) SLED profiles, normalized to the CO(2--1) transition.}}
        \label{fig:individual_comp_SLEDs}
\end{figure}

Using the classification {reported in Fig.~\ref{fig:fwhm_vcen}}, we computed velocity-integrated CO fluxes for each Gaussian component and, from these, we constructed CO SLEDs for all the quiescent and high-$v$/high-$\sigma$ components, separately. 
The results of this analysis are shown in Figure \ref{fig:individual_comp_SLEDs}, where the top panel shows in blue color the CO SLEDs of the `quiescent/non-outflowing' components (i.e., data points lying within the rectangle in Fig. \ref{fig:individual_comp_SLEDs}) and the bottom panel in green color shows the CO SLEDs of the high-$v$/high-$\sigma$ components (i.e., data points lying outside of the rectangle in Fig. \ref{fig:individual_comp_SLEDs}) that likely contain outflowing gas. 
The plots show a striking difference, where the CO ladders that are characteristic of the quiescent gas are significantly less excited than those produced by considering the high-v/high-sigma gas components.
This result, combined with the increase of line widths in higher-$J$ CO transitions shown in Figure~\ref{fig:all_linewidths_vs_j}, strongly suggests that the higher CO excitation seen in some ULIRGs is related to the appearance of prominent broad line wings, which likely contain massive molecular outflows in these sources. In other words, in those ULIRGs whose global CO SLEDs are characterised by a very high excitation, the source of such higher CO excitation is gas with a higher velocity shift and/or a higher velocity dispersion compared to disk gas, hence likely gas embedded in powerful, galaxy-wide outflows. Since our analysis is based on total (i.e. galaxy-integrated) line spectra, the trends shown in Fig.~\ref{fig:individual_comp_SLEDs} are remarkable, as they signify that the outflows {may be} affecting at the same time the global gas kinematics and the excitation of the molecular ISM in these sources.

\subsection{The origin of high CO excitation in ULIRGs}

The source of the highly excited CO gas in (U)LIRGs has been a long standing issue.
Previous studies of global CO SLEDs of ULIRGs used {\it Herschel} data, and so have been limited by the lack of spatial and/or spectral resolution, and by the use of heterogeneous datasets as the Herschel observations did not cover low-$J$ CO transitions. 
Nonetheless, these works showed that a classical PDR scenario is unable to account for the highly-excited mid-J CO SLEDs of ULIRGs (see Section~3 and references therein), and that other heating mechanisms are needed.
While low-$J$ CO lines are commonly optically thick and likely thermalized, populating high-$J$ CO lines usually requires non-thermal excitation mechanisms such as shocks, cosmic rays, or X-ray heating \citep{Esposito+22,Pereira-Santaella+13,Valentino+20}, and these mechanisms have been suggested in those case where a single PDR component fails to reproduce the observed CO SLED \citep[see e.g.,][]{Pereira-Santaella+14,Chen+23,Mingozzi+18,Pozzi+17}.
XDRs produced by the AGN are commonly invoked to account for higher-$J$ CO fluxes in ULIRGs. While there have been studies showing that the chemistry and physical conditions of the gas can be strongly impacted by AGN, either due to X-rays or shocks \citep[e.g.,][]{Aalto+11,Viti+14,Garcia-Burillo+14,Usero+04}, 
the consensus seems to be that emission associated with the AGN becomes significant only at $J \gtrsim 9$ CO lines \citep{Brusa+18,Lu+17,Esposito+22,Vallini+19}.

While we are not able to discern the exact origin of the CO excitation (e.g. radiative vs mechanical feedback) due to the lack of higher-$J$ CO constraints, our results, shown in Figures~\ref{fig:all_linewidths_vs_j} and ~\ref{fig:individual_comp_SLEDs}, challenge all scenarios for which the source of the high CO excitation in ULIRGs is a `classical' component of the ISM, such as PDRs formed around massive disc stars, or an XDR due to an AGN. 
Indeed, in such cases one would expect the highly-excited gas to reside both in the disc and in the outflow, but possibly to affect more the former than the latter, as it is the disc gas that is first exposed to the radiation coming from massive stars and AGNs.
Galactic-scale outflows develop from regions of massive star formation and/or from a central AGN, hence it is clear that radiation from these processes can leak through the outflows and so affect also the gas entrained at high velocities.
However, this would not explain why the bulk of the kpc-scale outflowing material would be significantly more affected than the disc, which is the interpretation we favor from our analysis of galaxy-integrated spectra.
Furthermore, because our analysis is based on galaxy-integrated single-dish spectroscopy, the observed trends cannot be driven by perturbed gas {confined within the} launching base of the outflow that is directly exposed to stellar and/or AGN radiation, as the emission from such component would be very compact, arising from scales of at most $\sim100$~pc, and so it would be diluted within our large beam. {Even in the case of gas heated to extreme temperatures by, for instance, shocks, in such compact regions, the areas sampled by our beams of $\sim10$~kpc are typically 10,000 times larger than the areas defined by the central $\sim100$~pc of these sources.}
Hence, we infer that the culprit for the globally high CO excitation of these ULIRGs must be gas that is significantly dynamically disturbed, and at the same time {occupies a substantial fraction of the beam} of our observations, so that its luminosity contributes significantly to the observed spectra in all CO transitions, producing a detectable signature.

{Previous} studies by \citet{Meijerink+13} and \citet{Papadopoulos+14} {focused} on the galaxy NGC~6240 argue that the high CO excitation is arising from galaxy-wide shocks. 
While this is not exactly the same mechanism, it demonstrates that shocks {can have indeed an impact on the shape of CO SLEDs and shocks are plausible to happen during the launch of molecular outflows.}
Thus, {the main question is to determine how much of the gas is affected by such outflow-related processes}, or, {alternatively,} how much of the source solid angle {is occupied by these components}, {which would determine to what degree} the global CO SLEDs will be impacted. To this end, it is then necessary to follow up with spatially resolved observations that allow for a proper characterization of the CO line luminosities in different galactic regions.
{In the context of previous results that point towards a mechanical feedback as the main culprit of high CO excitation in starburst galaxies, it is also worth mentioning the work of \cite{Harrington+21}. Their analysis uses a more realistic approach to CO SLED modelling that, once applied to a sample of high-redshift lensed dusty star forming galaxies, reveals the need for mechanical energy to be injected onto scales larger than kpc to drive the observed CO ladders.}

{In conclusion, although we are not able to pinpoint the exact physical mechanism at the origin of such link (e.g., heating due to outflow-driven shocks, large velocity gradients in the outflows, radiation or cosmic rays propagating mainly through the outflows without affecting the quiescent gas, etc)} our analysis has reliably shown there is a large fraction of the molecular gas in these ULIRGs that is both highly excited and significantly dynamically disturbed, likely contained within galaxy-scale outflows, while the dynamically quiescent gas is significantly less excited. {Although in our sample we did not detect a correlation between FIR colors and sources showing the most striking higher-$J$ CO line broadening (as discussed in Section~3.3), our proposed link between high global CO excitation and massive molecular outflows is not in contradiction with previous studies linking the high CO excitation with warmer FIR colors such as \cite{Pearson+16}. Indeed, one of the main mechanisms proposed for the acceleration of molecular outflows, especially in dusty galaxies hosting powerful AGN, is radiation pressure on dust \citep[see e.g.,][]{Costa+18b, Bieri+17, Ishibashi+23}, which thus could explain both properties of ULIRGs.}

{We note that molecular outflows show different properties in different classes of galaxies for what concerns their CO excitation, morphology/extent and energetics, probably reflecting a combination of (i) different ISM conditions of their host galaxies, (ii) different evolutionary stages/ages of the outflows themselves, and (iii) different driving mechanisms. This field is still in its infancy, and we are only now starting to address these issues. The literature reports both cases of low-excitation \citep[e.g., ESO320-G020, see][]{Pereira-Santaella+20} and high-excitation molecular outflows \citep[e.g., IC5063, see][]{Dasyra+16}. In this work, we are not proposing that all outflows (in ULIRGs or in other sources) have high excitation, but rather we are proposing a link between the exceptionally highly excited global CO conditions of ULIRGs (observed and noted since the early days of sub-millimeter astronomy) and the extremely high incidence \citep[$\sim70$\%, see e.g.,][]{Lamperti+22, Veilleux+13} of massive molecular outflows in these sources. The latter is indeed much higher than the incidence observed in normal main sequence star forming galaxies \citep[$\sim20-25$\%, see][]{Stuber+21}, which also show much lower global CO line excitation.}

\section{Summary and conclusions}\label{sec:conclusions}

We presented new, high S/N mid-$J$ CO line observations ($4 \leq J_{\rm up} \leq 7$) of 17 local ULIRGs obtained with the APEX telescope, using the nFLASH460 and SEPIA660 heterodyne receivers. We detected 32 out of the 33 targeted mid-$J$ CO lines, adding up to {74 out of 75} CO($J\leq7$) line detections. The only non-detection is due to an atmospheric line close to the tuning frequency, and most spectra have a high S/N that enable a detailed analysis of asymmetries and broad components of the spectral line profiles. For all these sources, we have very high S/N low-$J$ CO line observations (CO(1--0), CO(2--1), CO(3--2)), obtained by combining proprietary and archival APEX and ACA/ALMA data, which were presented in a previous paper \citep{MontoyaArroyave+23}.
Compared to previous multi-$J$ CO studies of local ULIRGs that combined heterogeneous ground-based low-$J$ CO data with high-frequency {\it Herschel} spectroscopy, our study can rely on a homogeneous, ground-based dataset, characterised by a high S/N and high spectral resolution. This survey was expressly designed to be sensitive to faint and diffuse components, such as those embedded in the massive molecular outflows that are almost ubiquitous in local ULIRGs, which  may have been missed by previous interferometric and/or low S/N analyses. On the other hand, our study is limited in that it lacks high-$J$ CO coverage, as there are no space-based facilities that can provide high-spectral resolution observations in the THz window. Our results can be summarised as follows:

\begin{itemize}
    \item Our sample exhibits a variety of CO SLED shapes, with peaks or plateaus emerging at different $J$ values, demonstrating that the ULIRG population can not be described by a single set of average physical conditions. 
    Some sources show weakly sub-thermally excited lines up to CO(4--3), and in particular, galaxy IRAS 10378+1109 shows almost thermalized emission up to CO(7--6).
    \item However, in general, these ULIRGs are characterised by rather extreme CO excitation conditions, with average CO line ratios that are higher than those measured in high-$z$ dusty star forming galaxies. We measure 
    average ratios of $\langle r_{41} \rangle = 0.54\pm0.03$ (based on 9 sources) and $\langle r_{61} \rangle = 0.170\pm0.007$ (based on 6 sources), where $r_{J1}\equiv L'_{{\rm CO}(J\rightarrow J-1)}/L'_{\rm CO(1-0)}$. The average $r_{51}$ and $r_{71}$ measurements, respectively $\sim0.42$ and $\sim0.58$, are biased towards higher values, since they are based only on two sources with high CO excitation.
    \item Based on the shape of their CO ladders, we classify our sample into lower-excitation sources, characterized by CO SLEDs that peak around $J_{\rm up}\sim3,4$, and higher-excitation sources, for which the CO SLEDs keep increasing or plateau up to the highest-$J$ CO line detected.
    \item A single component LVG modeling of the CO SLEDs of our sample, performed following the method developed by \cite{Yang+17}, delivers average volume densities $n_{\rm H_2} \approx 10^{2.0}-10^{5.8}$ cm$^{-3}$, with the majority (14 out of 17) of sources having moderate densities with $n_{\rm H_2}< 10^{3.3}$, and only three sources having high densities of $n_{\rm H_2}\gtrsim 10^{4.5}$.
    Gas kinetic temperatures vary from $T_{\rm kin}\approx16-795$ K, and CO column density per unit velocity gradient $N_{\rm CO}/{\rm d}v\approx10^{16.2}-10^{18.8}$ cm$^{-2}$ (km~s$^{-1}$)$^{-1}$.
    These values are, however, likely to be overestimated, especially for the higher-excitation ULIRGs, for which extremely high (unphysical) values are derived for $n_{\rm H_2}$ and/or $T_{\rm kin}$, in order to reproduce the observed CO fluxes, indicating the need of a second high excitation component. However, a two component LVG fit is highly degenerate with our data, because of the lack of  CO($J_{\rm up}>7$) lines, hence it is not presented in this work.
    \item We measured the line widths of each CO transition using both a Gaussian fit and a non parametric measurement, and analyzed their variations as a function of the CO transition ($J_{\rm up}$). We found a positive trend, statistically significant at the 4$\sigma$ level, of increasing CO line width with $J_{\rm up}$ of the transition, which appears to be driven by the higher-excitation sources in the sample.
    \item We then investigated further this trend, by focusing on the higher-excitation sources. Similar to the analysis already presented in \cite{MontoyaArroyave+23} for the low-$J$ CO lines, we performed a simultaneous multi-Gaussian spectral fit, where all Gaussian components are constrained to share the same central velocity and FWHM in all CO transitions available for each source, while only their amplitude is allowed to vary. We then analyzed separately the CO SLEDs obtained from the spectral components displaying low central velocities and/or small FWHM values, and those characterized by disturbed gas kinematics, i.e. with high velocity shifts and/or high FWHM values, which in these ULIRGs are likely dominated by massive molecular outflows. We found a clear difference in CO excitation between the CO ladders that are characteristic of dynamically quiescent gas are significantly less excited than those produced by considering the high velocity and/or high velocity dispersion gas.
    \item We find evidence, for the first time in a sample of local ULIRGs, that the higher CO excitation evidenced in some of these sources is related to the broadening of their CO lines. We suggest that the higher excitation gas is embedded in galaxy-scale molecular outflows, while it only contributes to a minor extent to the dynamically quiescent ISM of these sources.
\end{itemize}

Overall, we confirm the exceptional {\it global} molecular gas excitation of local ULIRGs, and we favor the interpretation that the origin of the higher excitation is associated to the large-scale outflows present in these sources, rather than the interpretation that the very hot components at the launching base of the outflows, or very hot turbulent gas confined in very small scales, are the drivers of the observed global physical conditions.
Moreover, our results challenge scenarios for which the high excitation phase of the molecular medium in ULIRGs is related to a classical PDR or XDR component residing in the dynamically quiescent disc gas, while they suggest that such phase can be found in high-velocity gas, likely containing outflows, and filling the beam of single-dish observations.

This study demonstrates the need to perform extensive high S/N line surveys with sensitive large-aperture single-dish telescopes that can capture also the faint, most diffuse and extended components of the molecular ISM of galaxies including the outflowing gas, and so it provides a key scientific driver for the development of the future Atacama Large Aperture Submillimetre Telescope (AtLAST\footnote{\texttt{https://www.atlast.uio.no}}, \citealt{Klaassen+20, Ramasawmy+22, Mroczkowski+23}) and its instrumentation. To perform the kind of analysis presented in this work, the latter needs to include high frequency ($\nu>350$~GHz), high spectral resolution instrumentation that can deliver stable baselines over a broad simultaneous bandwidth of at least $\sim4$~GHz, to enable the characterisation of broad and high-velocity emission line components of extragalactic sources.

\begin{acknowledgements}
    We want to give special thanks to the APEX and ESO staff for their efforts at continuing APEX operations and completing our projects despite the Covid19 pandemic.
    IMA would like to thank Davide Decataldo and Alice Schimek for useful discussions.
    {We thank the referee for valuable comments that helped improve the paper.}
    This project has received funding from the European Union’s Horizon 2020 research and innovation programme under grant agreement No 951815 (AtLAST).
    This publication is based on data acquired with the Atacama Pathfinder Experiment (APEX) under programme IDs 0104.B-0672, 0106.B-0674, 086.F-9321, 090.B-0404, 092.F-9325, 099.F-9709, 077.F-9300, and 084.F-9306. 
    APEX is a collaboration between the Max-Planck-Institut fur Radioastronomie, the European Southern Observatory, and the Onsala Space Observatory.
    This paper makes use of the following ALMA data: ADS/JAO.ALMA\#2016.1.00177.S, ADS/JAO.ALMA\#2013.1.00535.S, ADS/JAO.ALMA\#2016.1.00140.S, 
        ADS/JAO.ALMA\#2016.1.00177.S, ADS/JAO.ALMA\#2013.1.00180.S, ADS/JAO.ALMA\#2018.1.00503.S, ADS/JAO.ALMA\#2016.2.00006.S, 
        ADS/JAO.ALMA\#2017.1.01398.S, ADS/JAO.ALMA\#2017.1.00297.S, ADS/JAO.ALMA\#2013.1.00180.S, ADS/JAO.ALMA\#2018.1.00888.
    ALMA is a partnership of ESO (representing its member states), NSF (USA) and NINS (Japan), together with NRC (Canada), MOST and ASIAA (Taiwan), and KASI (Republic of Korea), in cooperation with the Republic of Chile. The Joint ALMA Observatory is operated by ESO, AUI/NRAO and NAOJ.
    This work is based on observations carried out with the IRAM Plateau de Bure Interferometer. IRAM is supported by INSU/CNRS (France), MPG (Germany), and IGN (Spain).
\end{acknowledgements}

\bibliography{biblio}
\bibliographystyle{aa}

\clearpage
\onecolumn

\begin{appendix}

\section{Measured fluxes and luminosities}
Tables \ref{table:fluxes} and \ref{table:luminosities} list the integrated line fluxes and luminosities, respectively, obtained from directly integrating the spectra within $v \in (-1000, 1000)$ \kms, after setting a threshold of $>1.5\sigma$ for each channel (see Sect.~\ref{sec:data_analysis}).

\begin{table*}[h]
\footnotesize
        \centering 
        \caption{CO flux$^{a}$ measurements for the ULIRGs in our sample.}\label{table:fluxes}
        \begin{tabular}{lcccccccccc}
                \hline
                \hline
                Source              &     CO(1--0)      &     CO(2--1)      &     CO(3--2)      &     CO(4--3)      &     CO(5--4)     &     CO(6--5)     &     CO(7--6)     \\
                \hline
                IRAS F01572+0009    & 7.2 $\pm$ 1.1     & 34.6 $\pm$ 2.3    & 56.7 $\pm$ 4.4    & 60.4 $\pm$ 11.5   & 62.8 $\pm$ 15.5  &                  &                  \\ 
                IRAS 03521+0028     &                   & 49.2 $\pm$ 1.8    & 79.7 $\pm$ 5.8    & 106.6 $\pm$ 13.2  &                  &                  &                  \\ 
                IRAS F05189-2524    & 43.8 $\pm$ 1.3    & 141.0 $\pm$ 6.2   & 205.4 $\pm$ 16.5  & 388.6 $\pm$ 51.4  &                  & 294.9 $\pm$ 16.0 &                  \\ 
                IRAS 06035-7102     & 43.2 $\pm$ 2.0    & 168.0 $\pm$ 6.4   & 327.4 $\pm$ 16.6  &                   &                  & 456.0 $\pm$ 31.4 &                  \\ 
                IRAS 06206-6315     &                   & 88.4 $\pm$ 5.2    & 130.2 $\pm$ 7.9   & 254.7 $\pm$ 25.8  &                  & 274.4 $\pm$ 36.5 &                  \\ 
                IRAS 08311-2459     &                   & 146.8 $\pm$ 5.6   & 257.6 $\pm$ 17.6  & 360.8 $\pm$ 26.4  &                  & 444.3 $\pm$ 47.4 &                  \\ 
                IRAS 09022-3615     &                   & 297.3 $\pm$ 8.1   & 465.0 $\pm$ 29.8  & 1010.3 $\pm$ 72.1 &                  & 704.6 $\pm$ 44.1 &                  \\ 
                IRAS 10378+1109     &                   & 47.4 $\pm$ 3.7    & 84.9 $\pm$ 6.5    & 127.1 $\pm$ 22.8  &                  &                  & 484.8 $\pm$ 50.1 \\ 
                IRAS F12112+0305    & 34.4 $\pm$ 1.1    & 190.9 $\pm$ 8.9   & 270.0 $\pm$ 14.2  & 350.6 $\pm$ 24.4  &                  & 230.1 $\pm$ 27.4 &                  \\
                IRAS F13451+1232    & 12.7 $\pm$ 0.7    & 44.4 $\pm$ 5.5    & 64.4 $\pm$ 5.7    & 43.8 $\pm$ 10.5   &                  &                  &                  \\ 
                IRAS 16090-0139     & 17.7 $\pm$ 1.2    & 102.2 $\pm$ 3.8   & 178.6 $\pm$ 13.3  & 267.3 $\pm$ 26.6  &                  &                  & 791.5 $\pm$ 41.9 \\ 
                IRAS 17208-0014     & 150.5 $\pm$ 2.0   & 593.3 $\pm$ 15.0  & 1041.2 $\pm$ 42.5 & 787.2 $\pm$ 80.0  &                  & 764.9 $\pm$ 53.8 &                  \\ 
                IRAS 19254-7245     &                   & 178.8 $\pm$ 7.2   &                   & 529.3 $\pm$ 56.8  &                  & 172.9 $\pm$ 29.1 &                  \\ 
                IRAS F20551-4250    & 68.4 $\pm$ 1.6    & 291.6 $\pm$ 8.2   & 416.1 $\pm$ 22.4  & 523.2 $\pm$ 46.2  &                  & 260.8 $\pm$ 34.5 &                  \\ 
                IRAS F22491-1808    &                   & 78.3 $\pm$ 5.0    & 130.8 $\pm$ 5.6   & 147.3 $\pm$ 18.0  &                  & 196.2 $\pm$ 21.1 &                  \\ 
                IRAS F23060+0505    & 11.0 $\pm$ 0.4    & 55.8 $\pm$ 3.2    & 75.0 $\pm$ 3.9    & 128.4 $\pm$ 15.2  & 133.9 $\pm$ 25.0 &                  & 126.7 $\pm$ 17.0 \\ 
                IRAS F23128-5919    & 51.7 $\pm$ 0.7    & 197.1 $\pm$ 5.2   & 329.9 $\pm$ 7.7   & 374.8 $\pm$ 57.2  &                  & 225.8 $\pm$ 24.8 &                  \\  
                \hline
        \end{tabular}

        \tablefoot{$^{a}$ Velocity-integrated flux densities in units of Jy \kms.}

\end{table*}

\begin{table*}[h]
\footnotesize
        \centering 
        \caption{CO line luminosities$^{a}$ computed for the ULIRGs in our sample.}\label{table:luminosities}
        \begin{tabular}{lcccccccccc}
                \hline
                \hline
                Source              &     CO(1--0)      &     CO(2--1)      &     CO(3--2)      &     CO(4--3)      &     CO(5--4)     &     CO(6--5)     &     CO(7--6)     \\
                \hline
                IRAS F01572+0009    & 9.9 $\pm$ 1.5     & 11.8 $\pm$ 0.8    & 8.6 $\pm$ 0.7     & 5.2 $\pm$ 1.0     & 3.4 $\pm$ 0.8    &                  &               \\ 
                IRAS 03521+0028     &                   & 14.5 $\pm$ 0.5    & 10.5 $\pm$ 0.8    & 7.9 $\pm$ 1.0     &                  &                  &               \\ 
                IRAS F05189-2524    & 3.9 $\pm$ 0.1     & 3.1 $\pm$ 0.1     & 2.0 $\pm$ 0.2     & 2.2 $\pm$ 0.3     &                  & 0.7 $\pm$ 0.0    &               \\ 
                IRAS 06035-7102     & 13.6 $\pm$ 0.6    & 13.2 $\pm$ 0.5    & 11.4 $\pm$ 0.6    &                   &                  & 4.0 $\pm$ 0.3    &               \\ 
                IRAS 06206-6315     &                   & 9.4 $\pm$ 0.6     & 6.2 $\pm$ 0.4     & 6.8 $\pm$ 0.7     &                  & 3.3 $\pm$ 0.4    &               \\ 
                IRAS 08311-2459     &                   & 18.6 $\pm$ 0.7    & 14.5 $\pm$ 1.0    & 11.4 $\pm$ 0.8    &                  & 6.3 $\pm$ 0.7    &               \\ 
                IRAS 09022-3615     &                   & 13.0 $\pm$ 0.4    & 9.1 $\pm$ 0.6     & 11.1 $\pm$ 0.8    &                  & 3.4 $\pm$ 0.2    &               \\ 
                IRAS 10378+1109     &                   & 11.2 $\pm$ 0.9    & 8.9 $\pm$ 0.7     & 7.5 $\pm$ 1.3     &                  &                  & 9.4 $\pm$ 1.0 \\ 
                IRAS F12112+0305    & 9.1 $\pm$ 0.3     & 12.7 $\pm$ 0.6    & 8.0 $\pm$ 0.4     & 5.8 $\pm$ 0.4     &                  & 1.7 $\pm$ 0.2    &               \\ 
                IRAS F13451+1232    & 9.5 $\pm$ 0.5     & 8.3 $\pm$ 1.0     & 5.4 $\pm$ 0.5     & 2.1 $\pm$ 0.5     &                  &                  &               \\ 
                IRAS 16090-0139     & 16.0 $\pm$ 1.1    & 23.2 $\pm$ 0.9    & 18.0 $\pm$ 1.3    & 15.2 $\pm$ 1.5    &                  &                  & 14.7 $\pm$ 0.8 \\ 
                IRAS 17208-0014     & 13.5 $\pm$ 0.2    & 13.3 $\pm$ 0.3    & 10.4 $\pm$ 0.4    & 4.8 $\pm$ 0.5     &                  & 1.9 $\pm$ 0.1    &               \\ 
                IRAS 19254-7245     &                   & 8.3 $\pm$ 0.3     &                   & 6.2 $\pm$ 0.7     &                  & 0.9 $\pm$ 0.2    &               \\ 
                IRAS F20551-4250    & 6.2 $\pm$ 0.1     & 6.6 $\pm$ 0.2     & 4.2 $\pm$ 0.2     & 3.0 $\pm$ 0.3     &                  & 0.7 $\pm$ 0.1    &               \\ 
                IRAS F22491-1808    &                   & 5.9 $\pm$ 0.4     & 4.4 $\pm$ 0.2     & 2.8 $\pm$ 0.3     &                  & 1.6 $\pm$ 0.2    &               \\ 
                IRAS F23060+0505    & 17.0 $\pm$ 0.6    & 21.6 $\pm$ 1.2    & 12.9 $\pm$ 0.7    & 12.4 $\pm$ 1.5    & 8.3 $\pm$ 1.5    &                  & 4.0 $\pm$ 0.5 \\ 
                IRAS F23128-5919    & 5.0 $\pm$ 0.1     & 4.8 $\pm$ 0.1     & 3.6 $\pm$ 0.1     & 2.3 $\pm$ 0.3     &                  & 0.6 $\pm$ 0.1    &               \\   
                \hline
        \end{tabular}

        \tablefoot{$^{a}$ Line luminosities (brightness temperatures) in units of $L^{\prime}_{\textrm{CO}}\times 10^{9}$ K \kms~pc$^2$.}

\end{table*}

\section{Reduced spectra and CO SLEDs}\label{app:spec_sleds}
Figures~\ref{fig:specF01572} to \ref{fig:specF23128} show, for each source of our sample, the final reduced CO line spectra obtained with APEX, the observed CO SLEDs, and the results of the LVG modeling performed with the code \texttt{RADEX\_EMCEE} developed by \cite{Yang+17}, as explained in Section~\ref{sec:lvg}.

Each spectrum is fitted a (multi)-Gaussian profile, following the description provided in Section \ref{sec:data_analysis}. The total fit is plotted with a solid line, and the individual Gaussian components are shown in dashed lines, all in the corresponding color for that transition.
For the higher-excitation sources we perform an additional simultaneous fit, as described in Section \ref{sec:mol_outflows}, and are also plotted on the spectra for the relevant sources in black lines, the total fit with a solid lines, and the individual components in dotted lines.

In each CO SLED plot, the dark purple solid line corresponds to the CO SLED solution from the maximum posterior probability, while the dashed orange line shows the CO SLED model from the median value of the marginal probability distribution of each parameter. The fainter purple lines correspond to a random set of 100 models within the $1\sigma$ quartile. The posterior probability distributions of each parameter are shown by the density-contour plots and the purple histograms, with contours increasing in steps of $0.5\sigma$.

\begin{figure}[htpb]
        \includegraphics[width=0.99\textwidth]{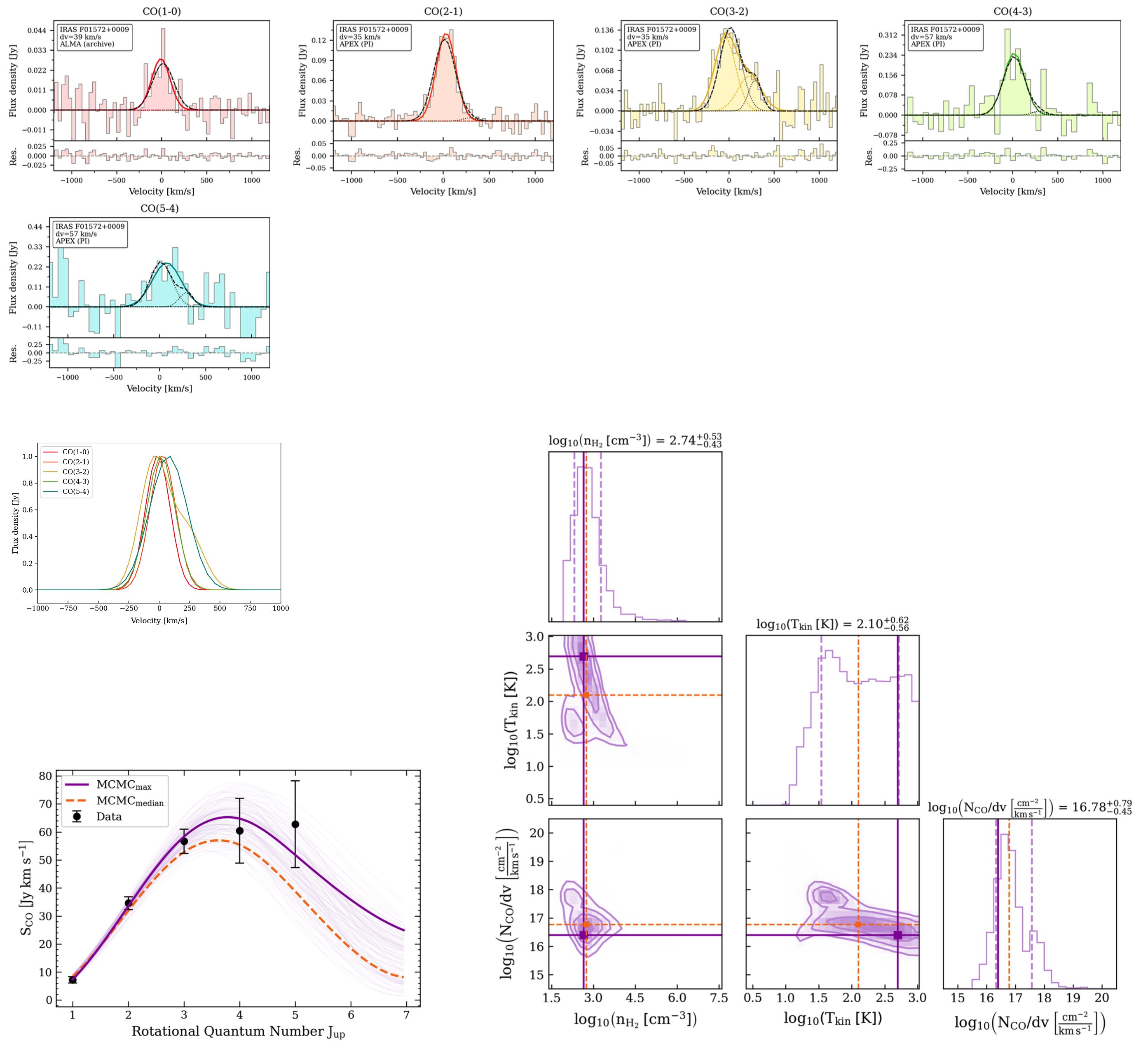}
        \caption{\textit{Top two rows:} Integrated and continuum-subtracted CO emission lines for galaxy IRAS F01572+0009. The spectral binning and telescope are reported on the top-left corner of each spectrum. The multi-Gaussian fit performed individually for each spectral line, is shown in corresponding colors to each CO transition (in a solid line is the total, and the individual components are shown in dashed lines).
        For the higher-excitation ULIRGs, the simultaneous multi-Gaussian fit is also overplotted on each spectrum in dark gray color (in a dashed line is the total, and the individual components are shown in dotted lines).
        \textit{Bottom panels:} On the top left are all the available spectral line profiles (fitted with multi-Gaussian profiles) normalized to their maximum value plotted together. This allows for better visualizing the variations in line widths for different CO rotational lines.
        On the bottom left is the CO SLED fitted with one gas component, where the maximum posterior probability is shown in a purple solid line, while the median value of the marginal probability distribution is shown with an orange dashed line.
        On the bottom right panel, and with the same colors, are the posterior probability distributions of the molecular gas density ($n_{\rm H_2}$), gas temperature ($T_{\rm kin}$) and CO column density per velocity ($N_{\rm CO}/{\rm dv}$) parameters for the single and two-component fit.}
 \label{fig:specF01572}
\end{figure}

\begin{figure}[htpb]
        \includegraphics[width=0.99\textwidth]{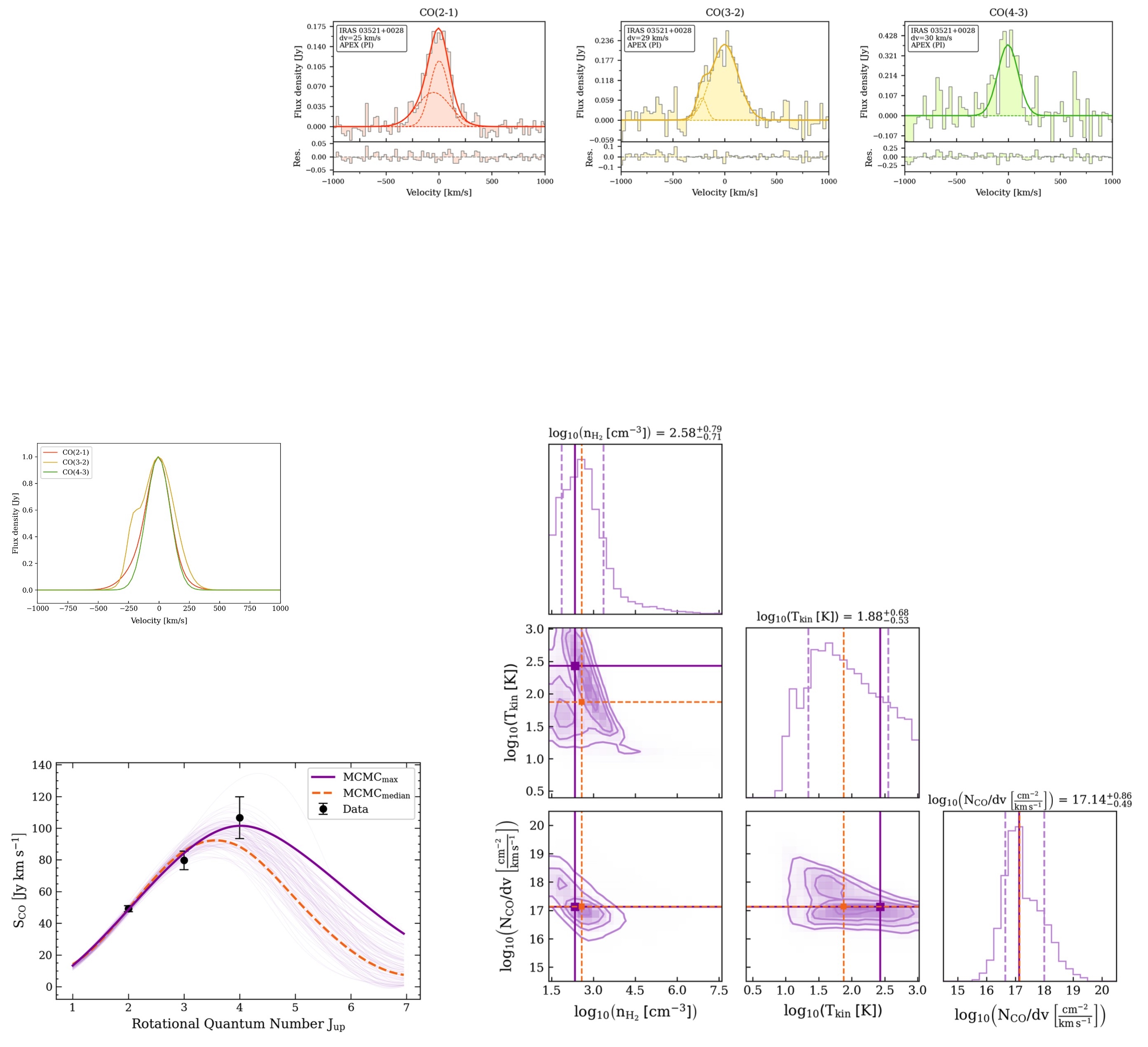}
        \caption{Continued from Fig. \ref{fig:specF01572} for source IRAS 03521+0028.}
 \label{fig:spec03521}
\end{figure}

\begin{figure}[htpb]
        \includegraphics[width=0.99\textwidth]{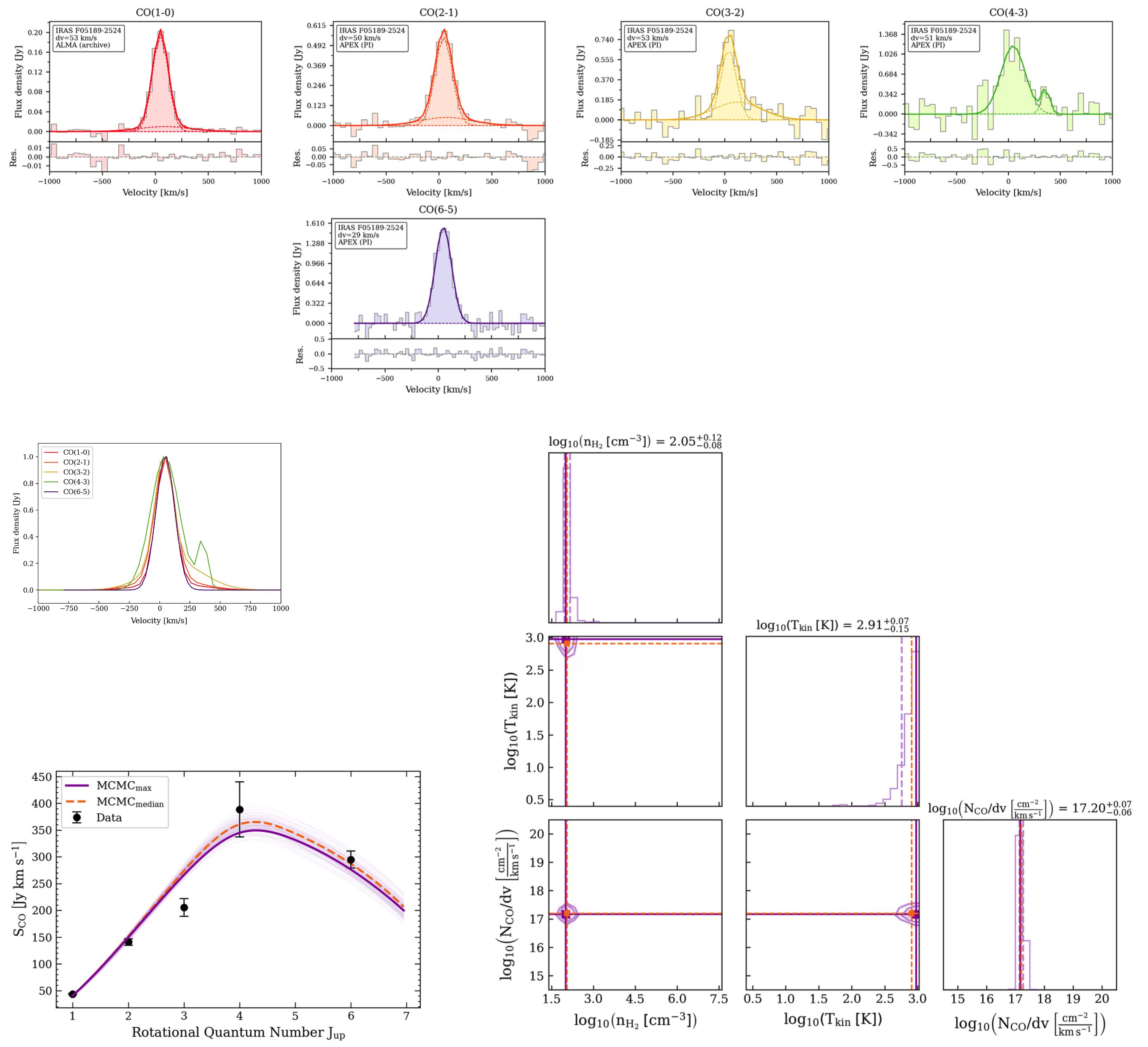}
        \caption{Continued from Fig. \ref{fig:specF01572} for source IRAS F05189-2524.}
 \label{fig:specF05189}
\end{figure}

\begin{figure}[htpb]
        \includegraphics[width=0.99\textwidth]{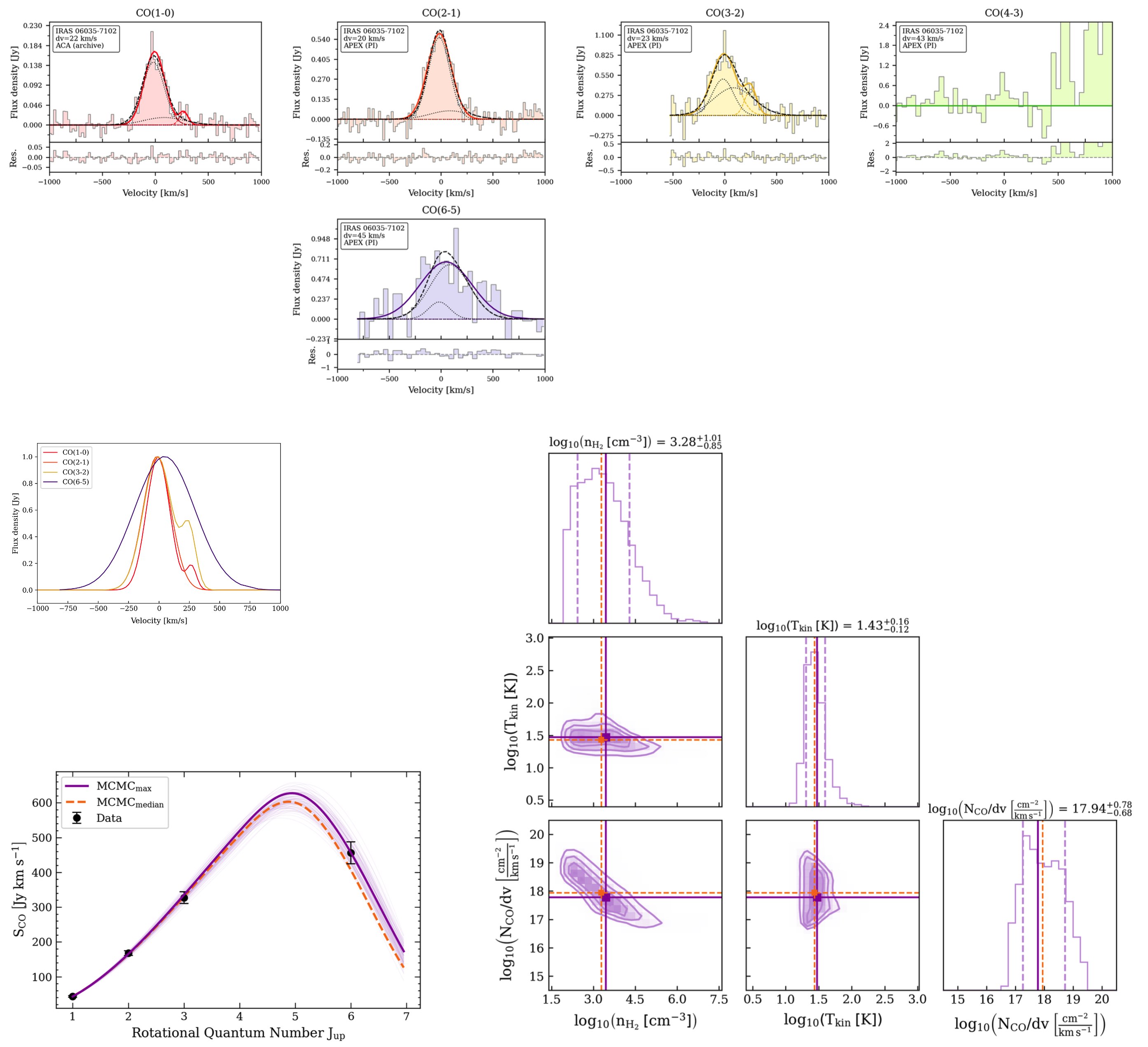}
        \caption{Continued from Fig. \ref{fig:specF01572} for source IRAS 06035-7102.}
 \label{fig:spec06035}
\end{figure}

\begin{figure}[htpb]
        \includegraphics[width=0.99\textwidth]{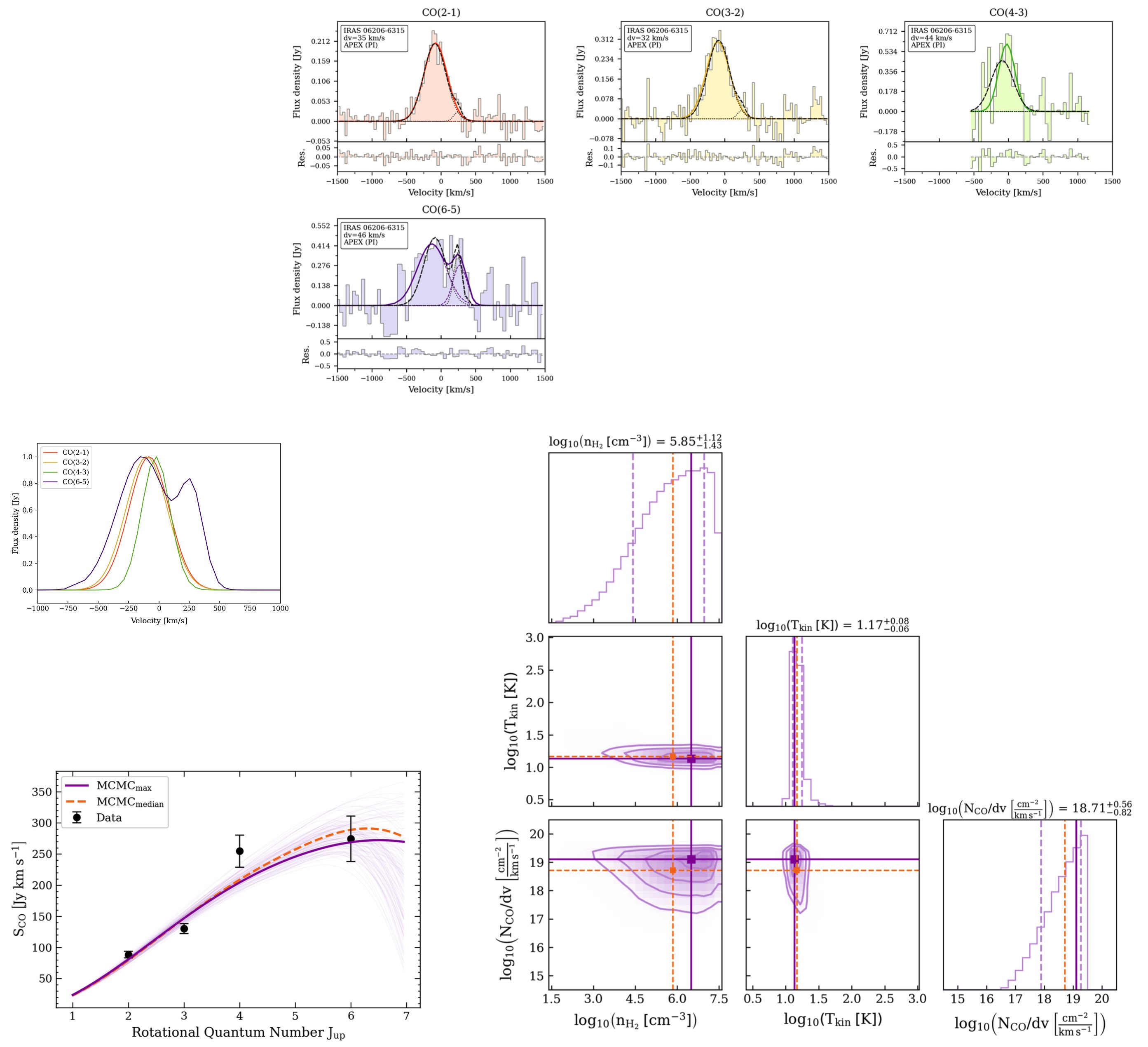}
        \caption{Continued from Fig. \ref{fig:specF01572} for source IRAS 06206-6315.}
 \label{fig:spec06206}
\end{figure}

\begin{figure}[htpb]
        \includegraphics[width=0.99\textwidth]{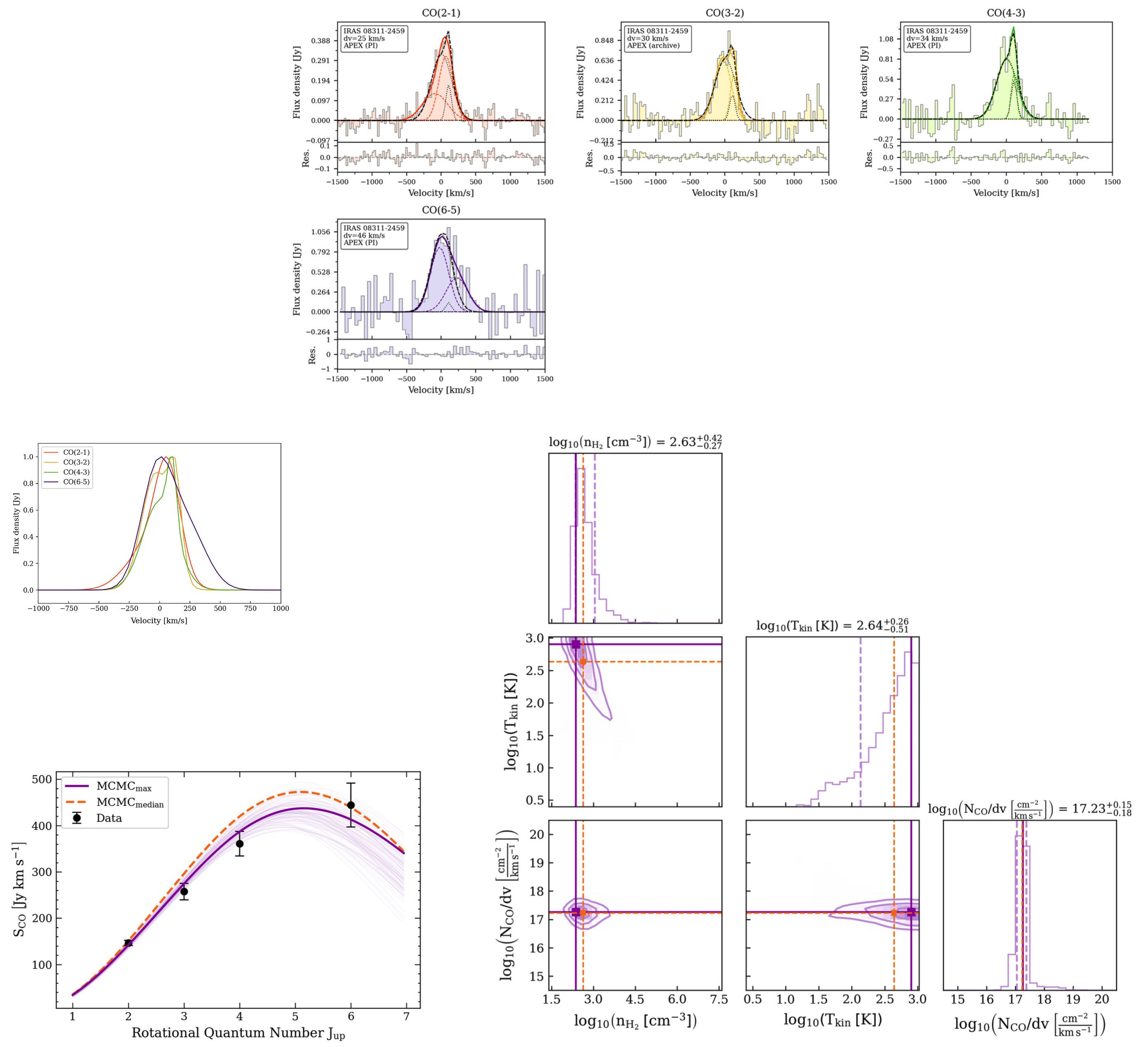}
        \caption{Continued from Fig. \ref{fig:specF01572} for source IRAS 08311-2459.}
 \label{fig:spec08311}
\end{figure}

\begin{figure}[htpb]
        \includegraphics[width=0.99\textwidth]{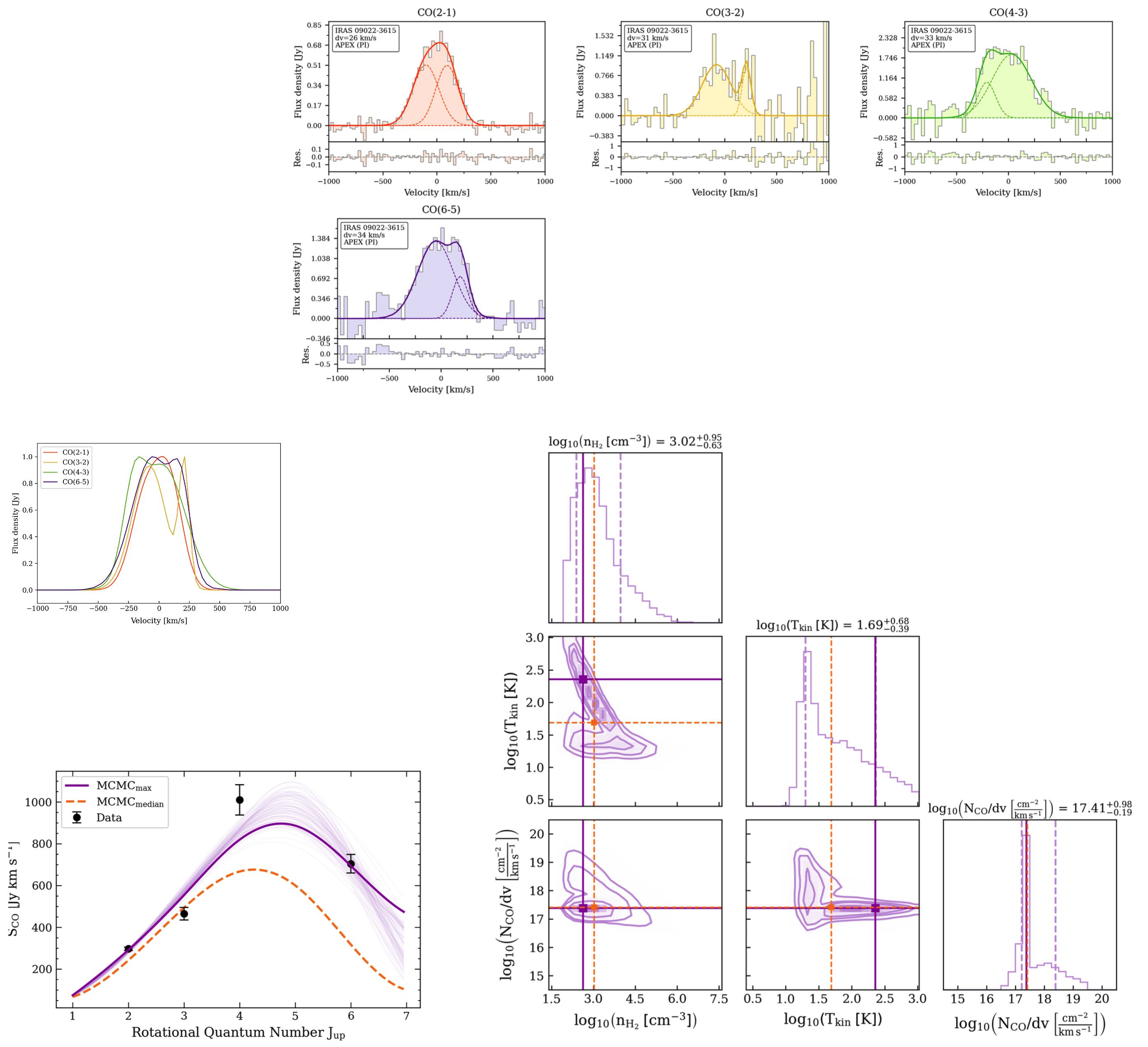}
        \caption{Continued from Fig. \ref{fig:specF01572} for source IRAS 09022-3615.}
 \label{fig:spec09022}
\end{figure}

\begin{figure}[htpb]
        \includegraphics[width=0.99\textwidth]{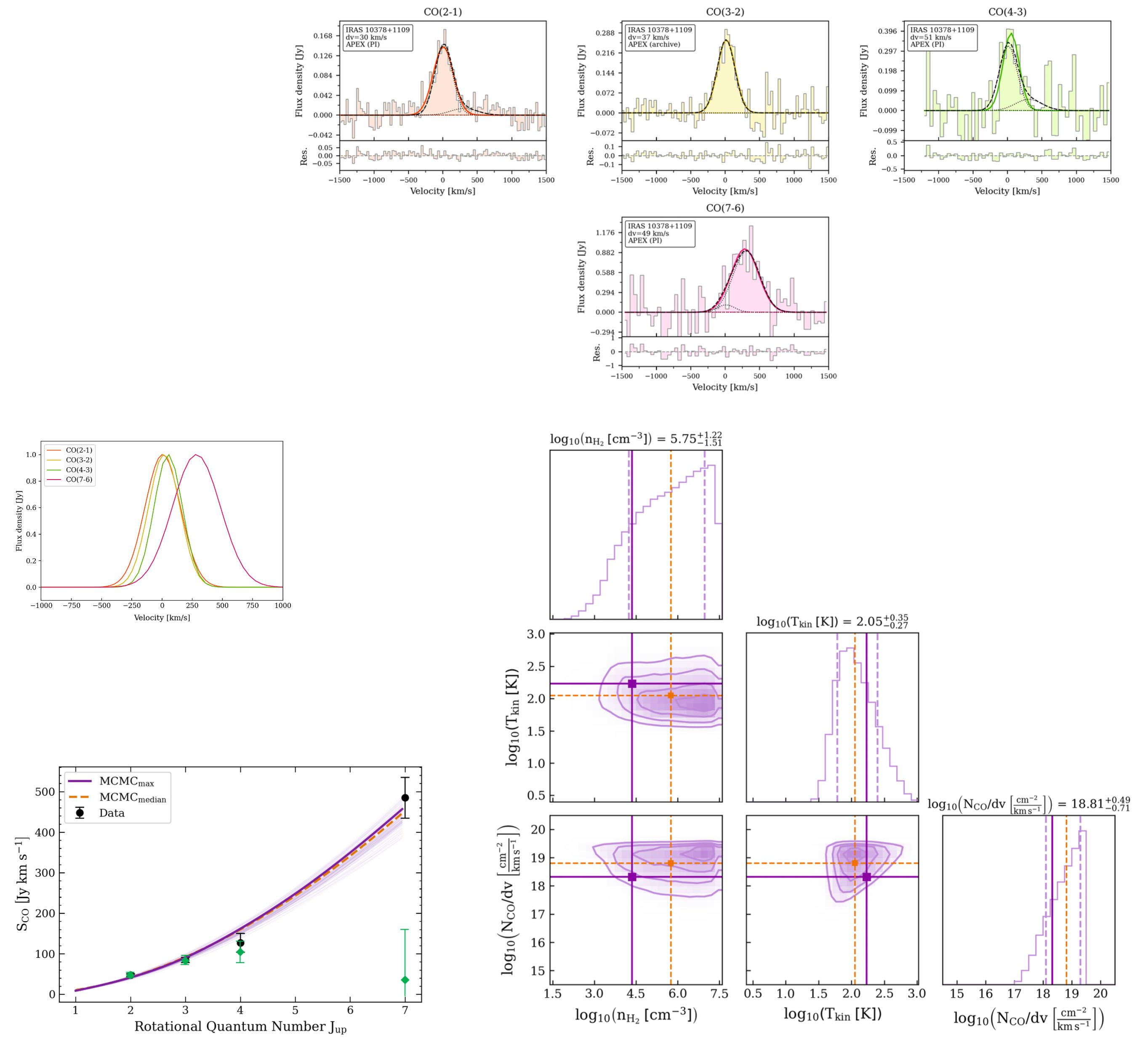}
        \caption{Continued from Fig. \ref{fig:specF01572} for source IRAS 10378+1109. {The green data points in the bottom-left panel showing the CO SLED correspond to the fluxes of the systemic velocity components (with central velocity $v_{\rm cen}\approx12$ \kms), resulting from the simultaneous fit. If considering only this systemic component, the CO SLED looks much less extreme, and in agreement with sub-thermal excitation conditions. Nonetheless, since the data analyzed in this work does not allow us to determine the origin of the velocity offset observed in the CO(7--6) spectrum, and for consistency with the other sources, we use the modeling results obtained by considering the CO SLED that uses the total velocity-integrated CO fluxes.}}
 \label{fig:spec10378}
\end{figure}

\begin{figure}[htpb]
        \includegraphics[width=0.99\textwidth]{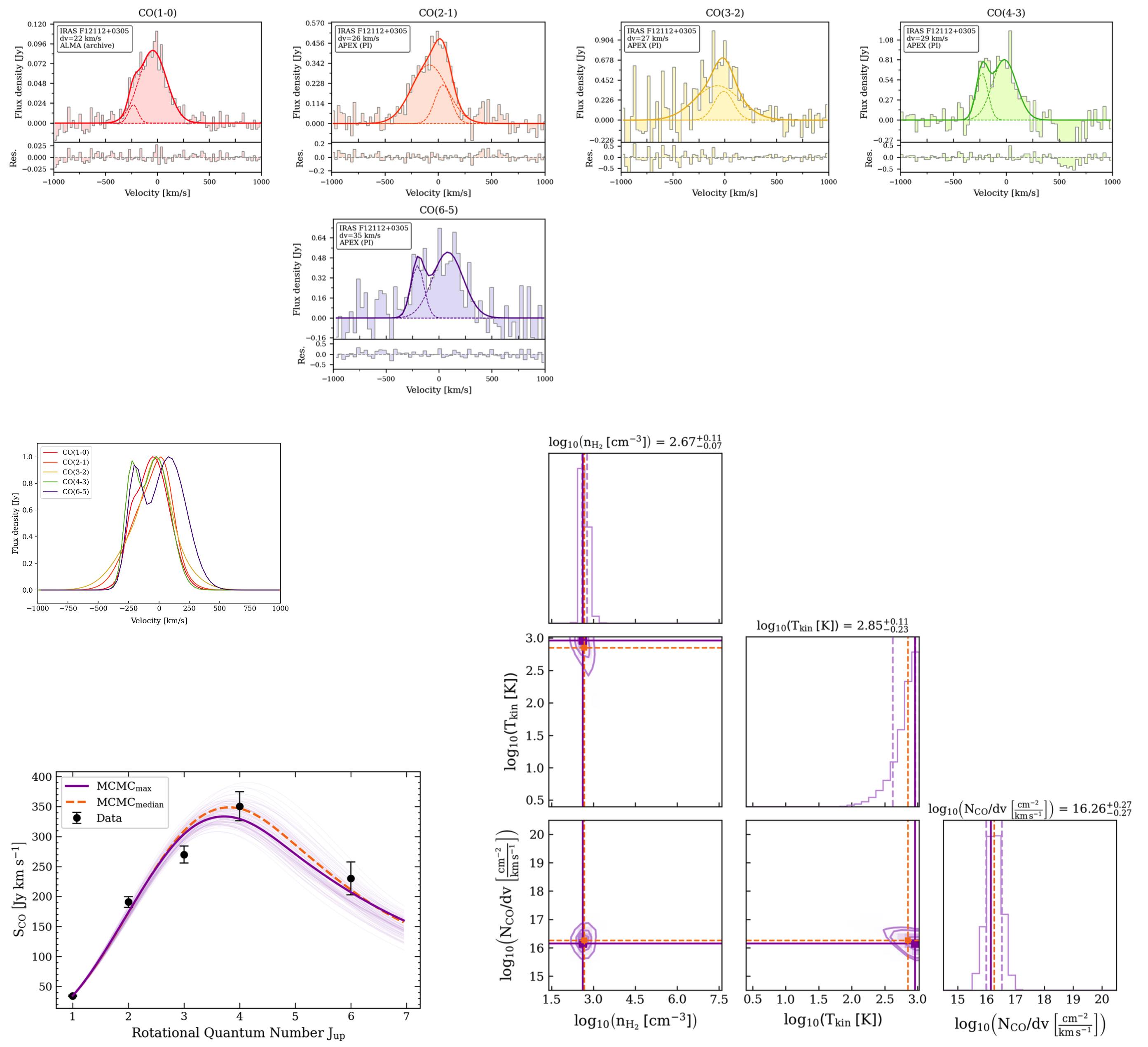}
        \caption{Continued from Fig. \ref{fig:specF01572} for source IRAS F12112+0305.}
 \label{fig:specF12112}
\end{figure}

\begin{figure}[htpb]
        \includegraphics[width=0.99\textwidth]{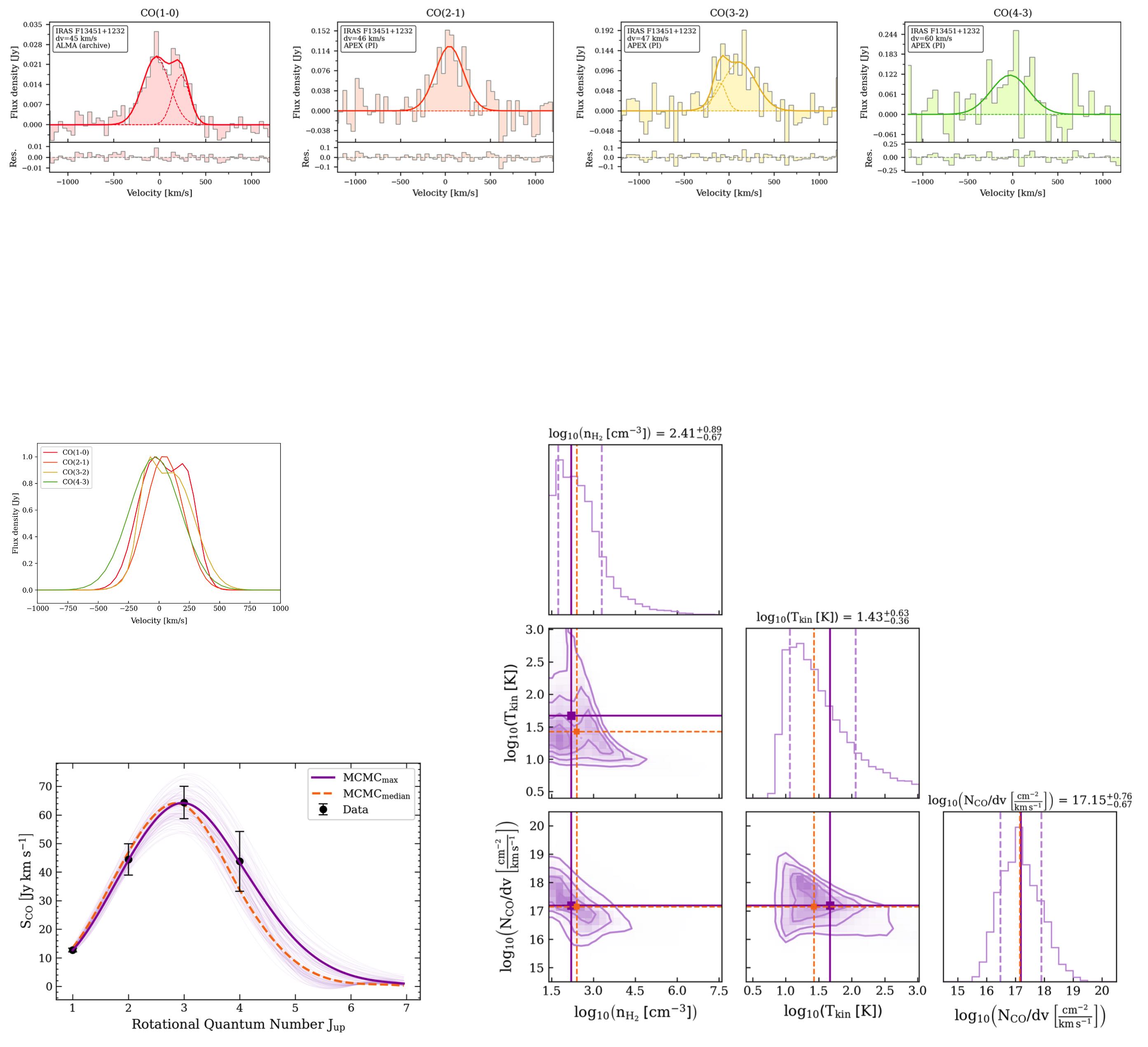}
        \caption{Continued from Fig. \ref{fig:specF01572} for source IRAS F13451+1232.}
 \label{fig:specF13451}
\end{figure}

\begin{figure}[htpb]
        \includegraphics[width=0.99\textwidth]{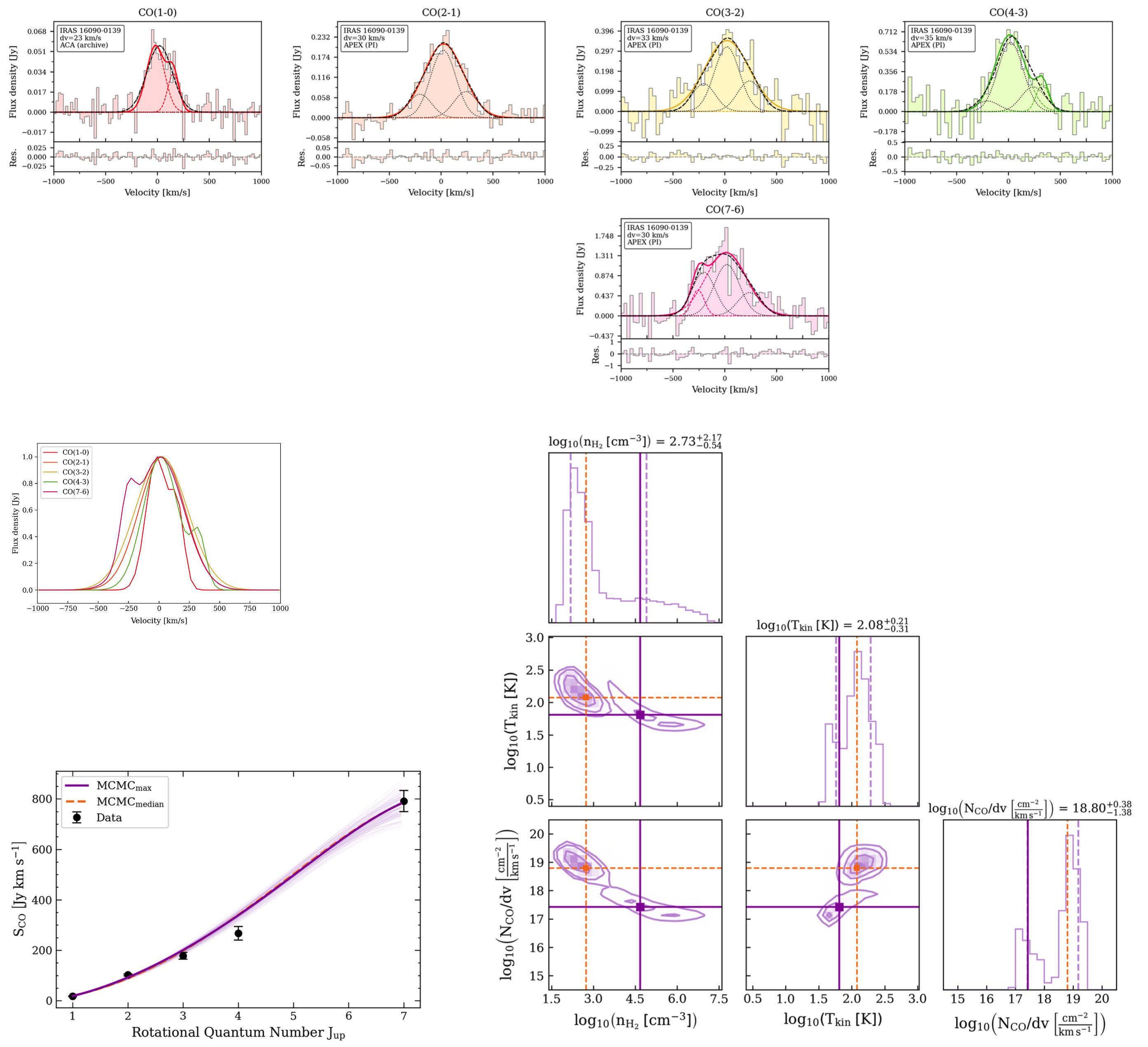}
        \caption{Continued from Fig. \ref{fig:specF01572} for source IRAS 16090-0139.}
 \label{fig:spec16090}
\end{figure}

\begin{figure}[htpb]
        \includegraphics[width=0.99\textwidth]{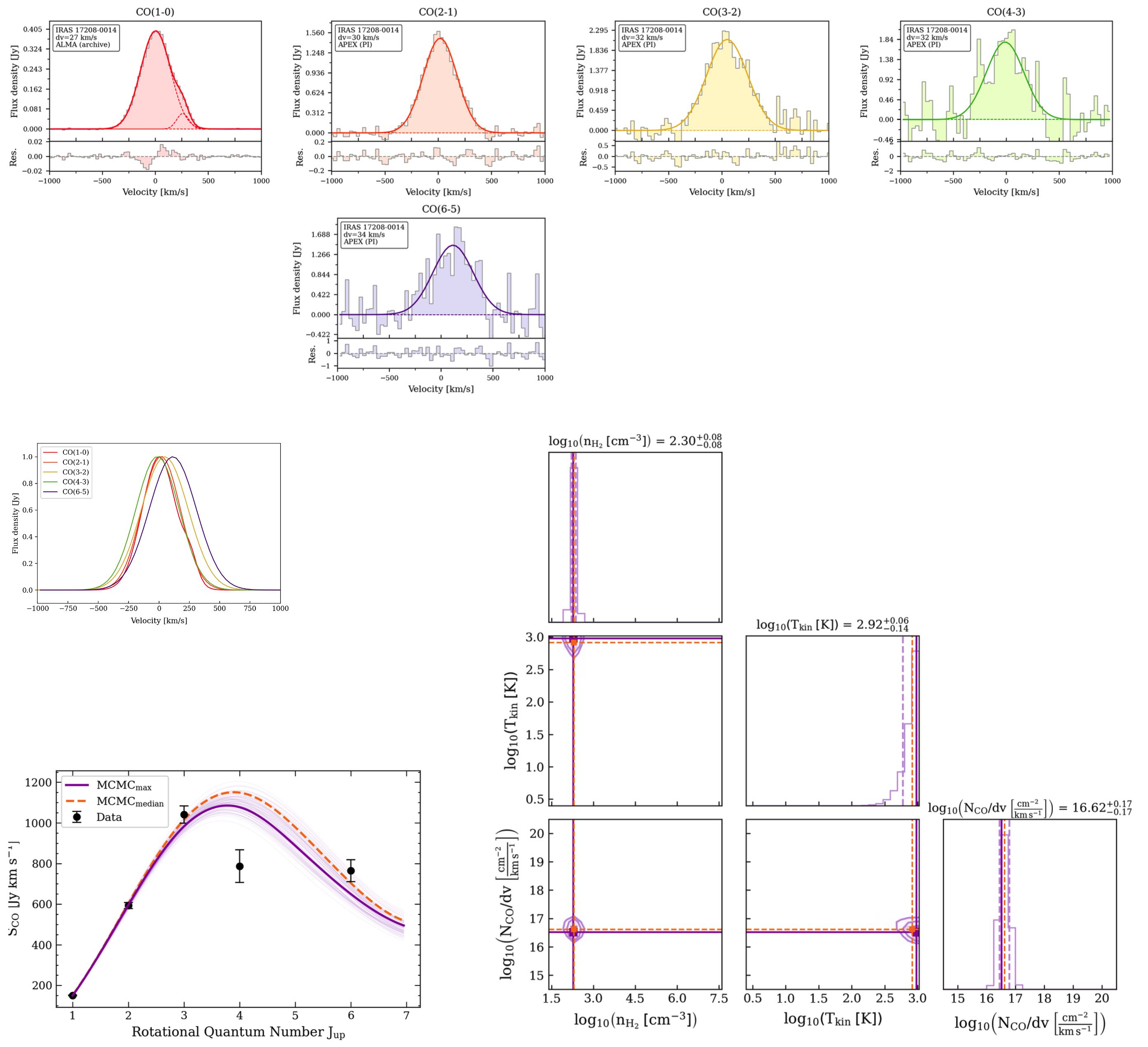}
        \caption{Continued from Fig. \ref{fig:specF01572} for source IRAS 17208-0014.}
 \label{fig:spec17208}
\end{figure}

\begin{figure}[htpb]
        \includegraphics[width=0.99\textwidth]{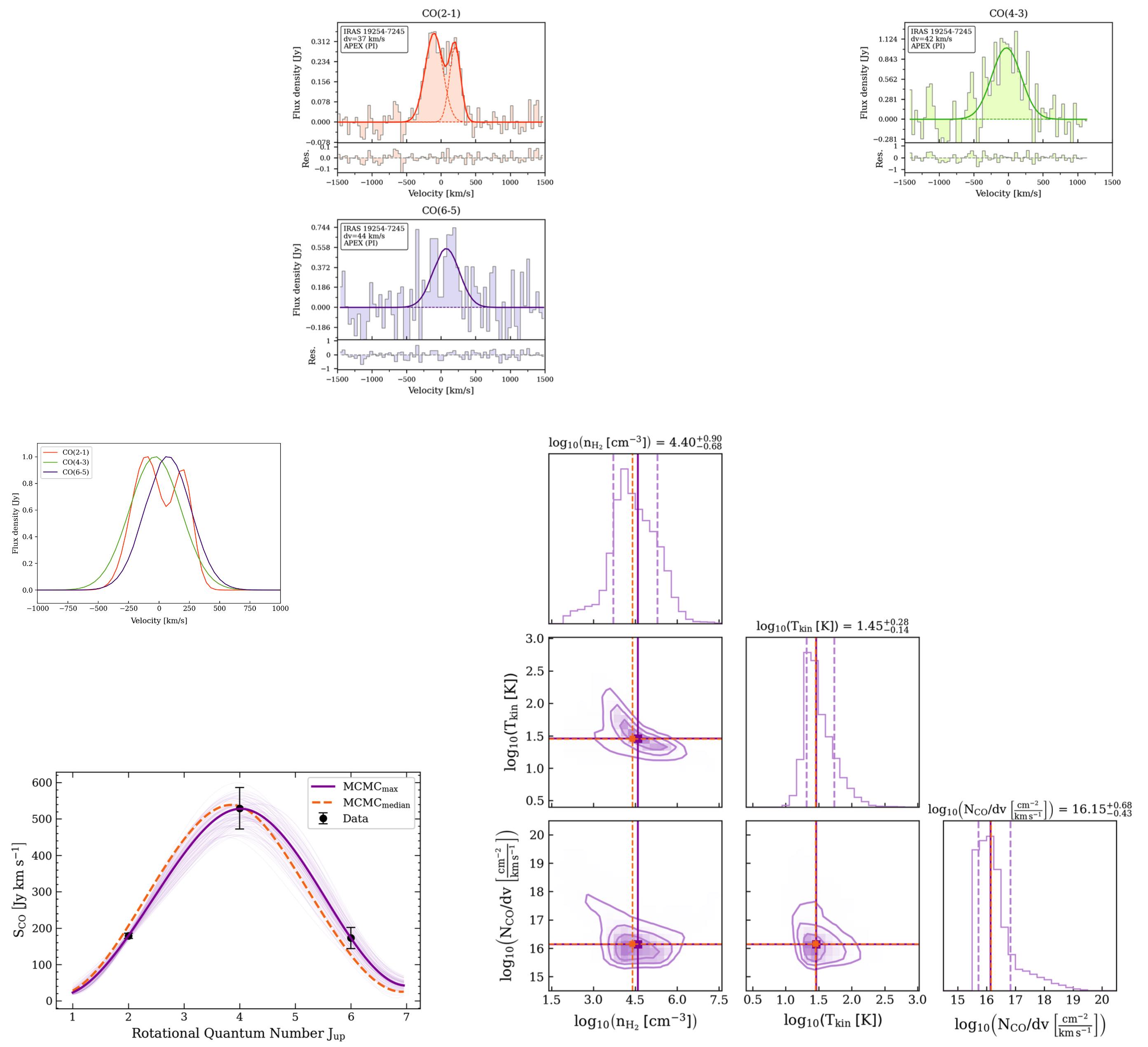}
        \caption{Continued from Fig. \ref{fig:specF01572} for source IRAS 19254-7245.}
 \label{fig:spec19254}
\end{figure}

\begin{figure}[htpb]
        \includegraphics[width=0.99\textwidth]{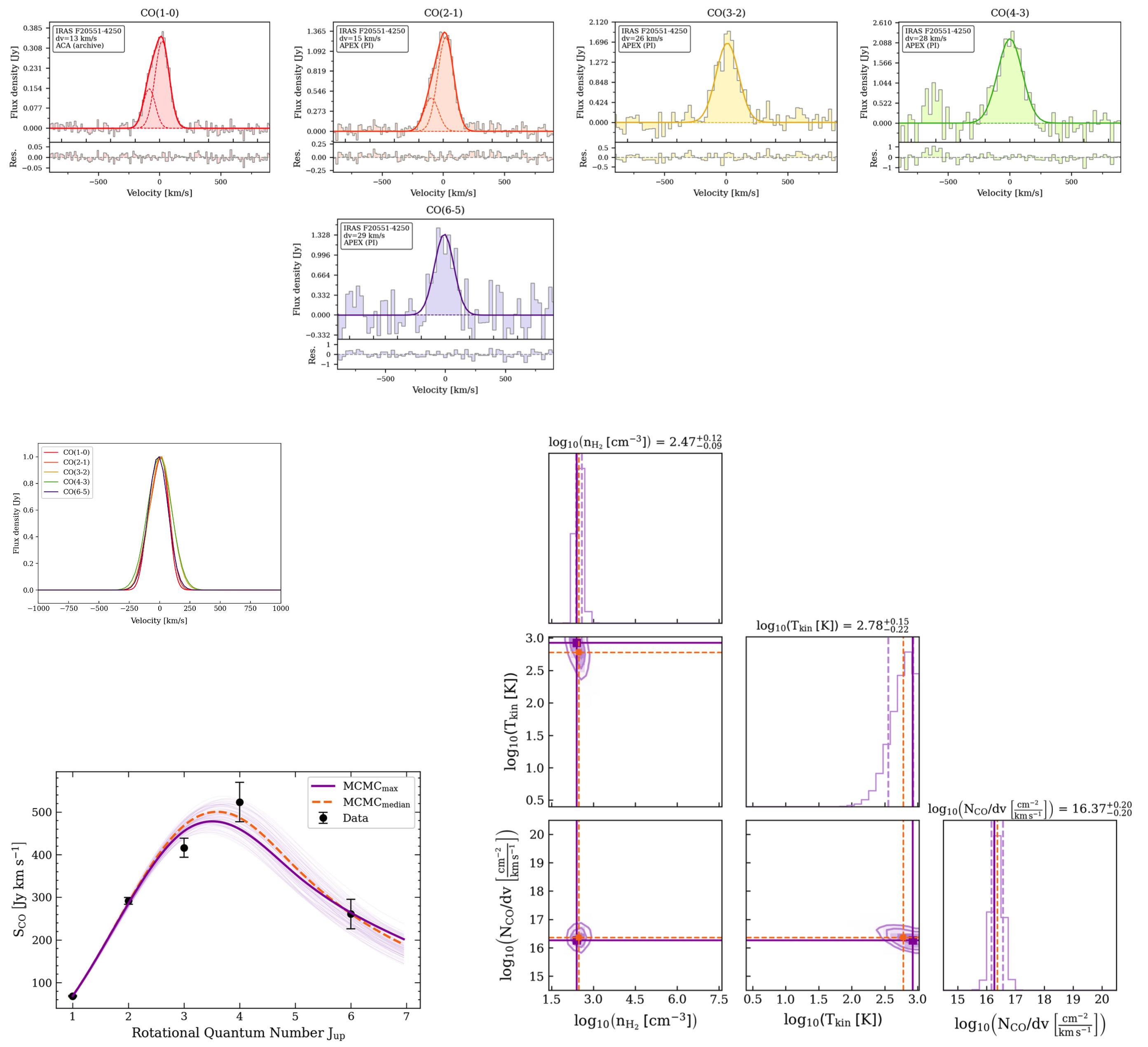}
        \caption{Continued from Fig. \ref{fig:specF01572} for source IRAS F20551-4250.}
 \label{fig:specF20551}
\end{figure}

\begin{figure}[htpb]
        \includegraphics[width=0.99\textwidth]{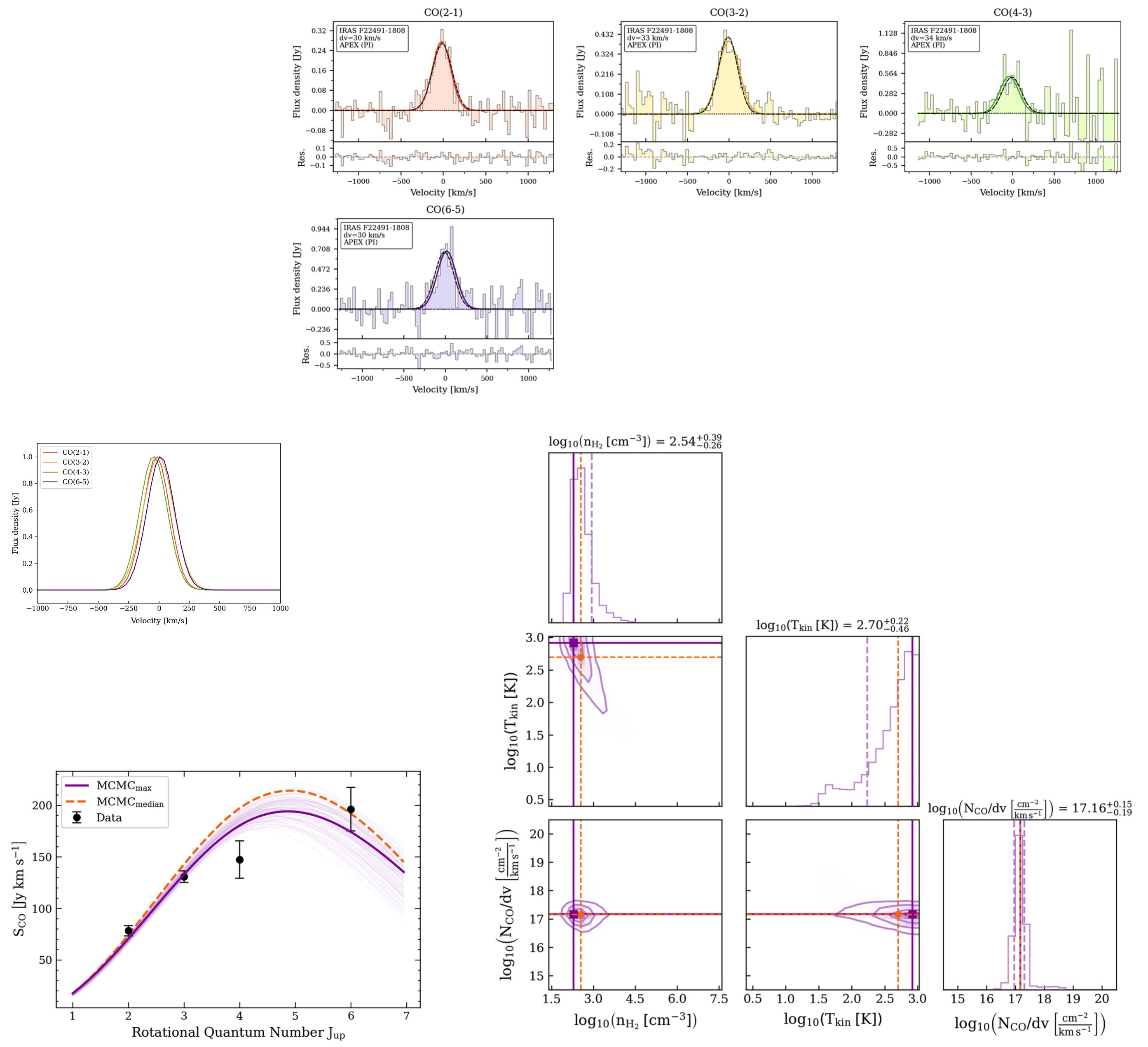}
        \caption{Continued from Fig. \ref{fig:specF01572} for source IRAS F22491-1808.}
 \label{fig:specF22491}
\end{figure}

\begin{figure}[htpb]
        \includegraphics[width=0.99\textwidth]{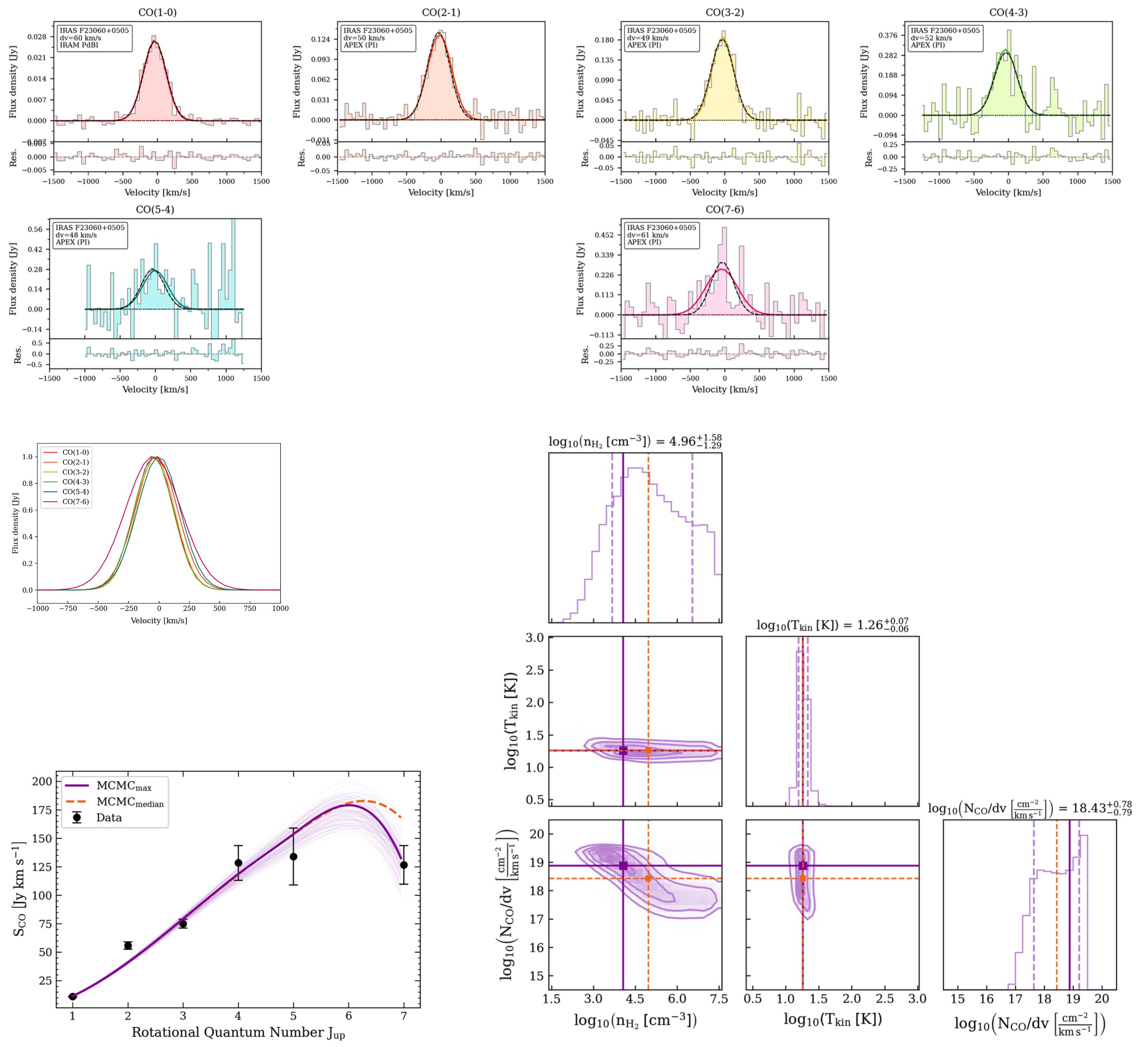}
        \caption{Continued from Fig. \ref{fig:specF01572} for source IRAS F23060+0505.}
 \label{fig:specF23060}
\end{figure}

\begin{figure}[htpb]
        \includegraphics[width=0.99\textwidth]{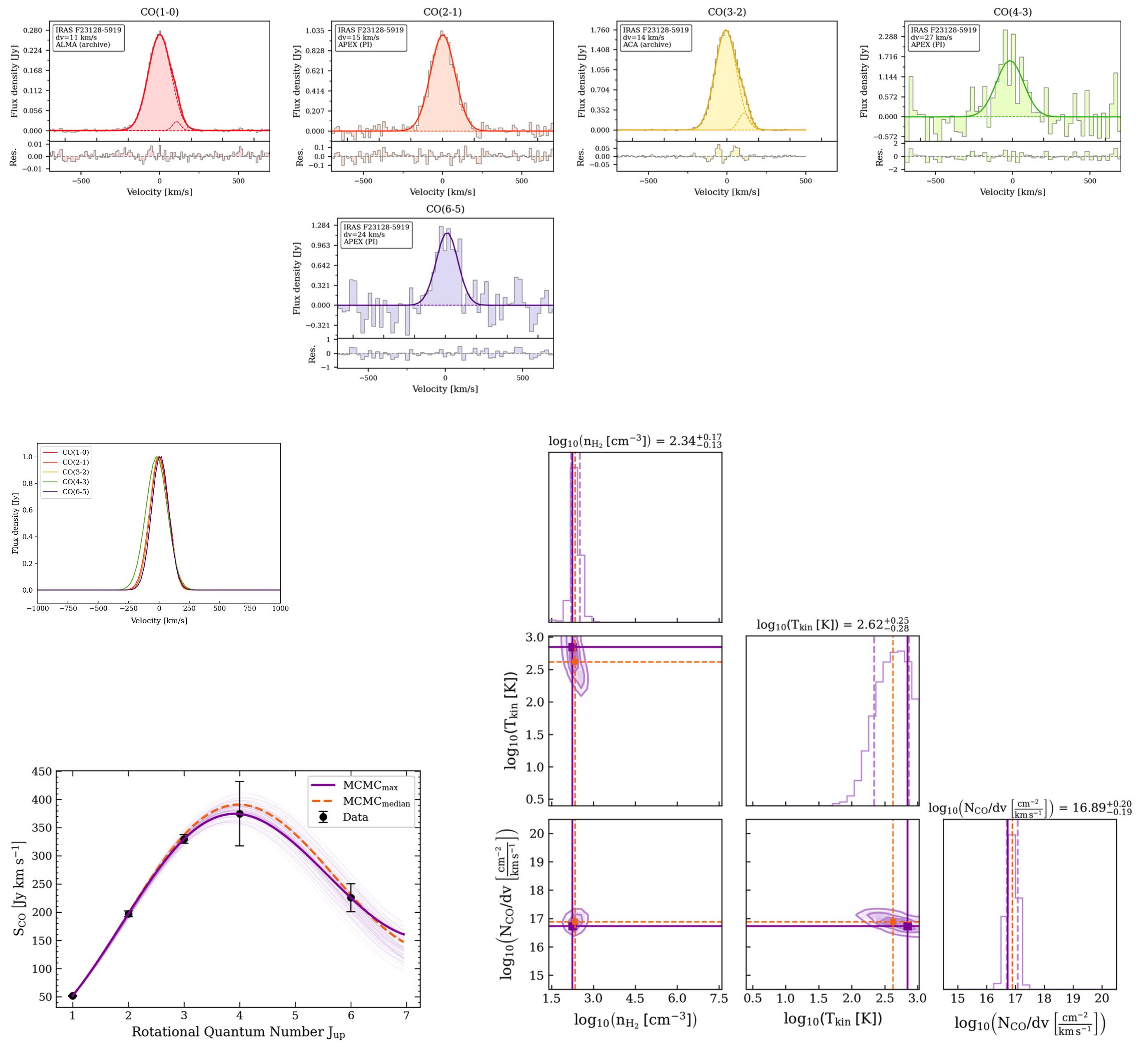}
        \caption{Continued from Fig. \ref{fig:specF01572} for source IRAS F23128-5919.}
 \label{fig:specF23128}
\end{figure}

\end{appendix}

\end{document}